\documentclass{IEEEtran}
\usepackage{cite}
\usepackage{amsmath,amssymb,amsfonts}
\usepackage{algorithmic}
\usepackage{graphicx}
\usepackage{textcomp}

\usepackage{hyperref}
\usepackage{physics}

\usepackage{bbm}
\usepackage{empheq}
\usepackage{braket}
\usepackage{booktabs}
\usepackage{makecell}
\usepackage{balance}

\usepackage[T1]{fontenc}

\usepackage{cleveref}
\crefname{equation}{}{}
\Crefname{Equation}{Equation}{Equations}
\crefname{table}{Table}{Tables}
\Crefname{table}{Table}{Tables}
\crefname{figure}{Fig.}{Fig.}
\Crefname{figure}{Fig.}{Fig.}

\def\ttr{t_{\textnormal{trunc}}}
\def\Zf{Z_{\textnormal{f}}}
\def\Zs{Z_{\textnormal{s}}}
\def\pdist{p_{\textnormal{dist}}}

\def\Wsuc{W'_{\textnormal{s}}}
\def\tA{t_{\textnormal{A}}}
\def\tB{t_{\textnormal{B}}}
\def\TA{T_{\textnormal{A}}}
\def\TB{T_{\textnormal{B}}}
\def\wA{w_{\textnormal{A}}}
\def\WA{W_{\textnormal{A}}}
\def\wB{w_{\textnormal{B}}}
\def\WB{W_{\textnormal{B}}}
\def\Ps{P_{\textnormal{s}}}
\def\Pf{P_{\textnormal{f}}}

\def\pto{p_{\textnormal{cut}}}
\def\Wprep{W_{\textnormal{s}}}
\def\tsuc{t_{\textnormal{suc}}}

\def\Yt{Y_{\textnormal{cut}}}

\def\Tout{T_{\textnormal{out}}}
\def\Wout{W_{\textnormal{out}}}

\def\SWAP{{\sc swap}}
\def\DIST{{\sc dist}}
\def\GEN{{\sc gen}}

\def\CUTOFF{{\sc cut-off}}
\def\pgen{p_{\textnormal{gen}}}
\def\pswap{p_{\textnormal{swap}}}
\def\tcoh{t_{\textnormal{coh}}}
\def\wout{w_{\textnormal{out}}}
\def\PUNIT{{\bf \sc protocol-unit}}

\def\cutoffdifference{\mbox{{\sc dif-time}-\CUTOFF}}
\def\cutoffmaximum{\mbox{{\sc max-time}-\CUTOFF}}
\def\cutofffidelity{\mbox{{\sc fidelity}-\CUTOFF}}

\newcommand{\Conv}{\mathop{\scalebox{3.0}{\raisebox{-0.2ex}{$*$}}}}%
\newcommand{\conv}{*}

\def\BibTeX{{\rm B\kern-.05em{\sc i\kern-.025em b}\kern-.08em
    T\kern-.1667em\lower.7ex\hbox{E}\kern-.125emX}}

\begin{document}
\title{Efficient Optimization of Cut-offs in Quantum Repeater Chains}

\author{\IEEEauthorblockN{Boxi Li\IEEEauthorrefmark{1}\IEEEauthorrefmark{2}\IEEEauthorrefmark{3},
Tim Coopmans\IEEEauthorrefmark{2}, and
David Elkouss\IEEEauthorrefmark{2}}

\IEEEauthorblockA{\IEEEauthorrefmark{1}ETH Z\"{u}rich, R\"{a}mistrasse 101, 8092,
Z\"{u}rich, Switzerland}

\IEEEauthorblockA{\IEEEauthorrefmark{2}QuTech, Delft University of Technology, Lorentzweg 1, 2628 CJ Delft, The Netherlands}

\IEEEauthorblockA{\IEEEauthorrefmark{3}Peter Gr\"unberg Institute - Quantum Control (PGI-8), Forschungszentrum J\"ulich GmbH, D-52425 J\"ulich, Germany}%
\thanks{This work has been submitted to the IEEE for possible publication. Copyright may be transferred without notice, after which this version may no longer be accessible.
Corresponding author: David Elkouss (email: D.ElkoussCoronas@tudelft.nl).}}

\maketitle


\begin{abstract}
    Quantum communication enables the implementation of tasks that are unachievable with classical resources.
However, losses on the communication channel preclude the direct long-distance transmission of quantum information in many relevant scenarios. In principle quantum repeaters allow one to overcome losses. However, realistic hardware parameters make long-distance quantum communication a challenge in practice. 
For instance, in many protocols an entangled pair is generated that needs to wait in quantum memory until the generation of an additional pair. During this waiting time the first pair decoheres, impacting the quality of the final entanglement produced. At the cost of a lower rate, this effect can be mitigated by imposing a cut-off condition. For instance, a maximum storage time for entanglement after which it is discarded. In this work, we optimize the cut-offs for quantum repeater chains. First, we develop an algorithm for computing the probability distribution of the waiting time and fidelity of entanglement produced by repeater chain protocols which include a cut-off.
Then, we use the algorithm to optimize cut-offs in order to maximize secret-key rate between the end nodes of the repeater chain.
We find that the use of the optimal cut-off extends the parameter regime for which secret key can be generated and moreover significantly increases the secret-key rate for a large range of parameters.

\end{abstract}

\section{Introduction}
The realization of a quantum internet~\cite{kimble2008quantum} will allow any two parties on earth to implement tasks that are impossible with its classical counterpart \cite{wehner2018quantum}.
Quantum communication schemes rely on the transmission of quantum information, which in practice is precluded over long distances due to loss in the communication channel (usually glass fiber or free space).
This problem can be overcome by dividing the distance between the sender and receiver of the quantum information into smaller segments, which are connected by intermediate nodes called quantum repeaters \cite{briegel1998quantum}.

Most repeater schemes require quantum memories~\cite{munro2015inside,muralidharan2016optimal}. Moreover, in many protocols an entangled pair is generated that needs to wait in a quantum memory until the generation of an additional pair. During this waiting time the first pair decoheres, reducing the quality of the final entanglement produced. At the cost of a lower rate, this effect can be mitigated by imposing a cut-off condition. For instance, a maximum storage time for entanglement after which it is discarded~\cite{collins2007multiplexed}. 

Cut-offs have been considered for entanglement generation in different contexts \cite{collins2007multiplexed,praxmeyer,kalb2017entanglement,rozpedek2018near-term,rozpedek2018parameters,santra2018quantum,chakraborty2019distributed,loock2020extending,schmidt2020memory,khatri2019practical,shchukin2019waiting,wu2020nearterm}. Notably, they play a key role for generating entanglement already in multi-pair experiments between adjacent nodes~\cite{kalb2017entanglement}. They also promise to be helpful in near-term quantum repeater experiments~\cite{rozpedek2018near-term,rozpedek2018parameters,schmidt2020memory}.
In the multi-repeater case, it is possible to obtain analytical expressions for the waiting time for general families of protocols \cite{khatri2019practical,shchukin2019waiting}, though in general it appears challenging to extend those methods to characterize the quality of the states produced.
Santra et al.~\cite{santra2018quantum} analytically optimized the distillable entanglement for a restricted class of quantum repeater schemes.

In this work, we first characterize the performance of a very general class of repeater schemes including cut-offs, probabilistic swapping, distillation and memory decoherence.
We sidestep the challenge of analytical characterization by computing the probability distribution of the waiting time and fidelity of the first generated entangled pair between the repeater's end nodes.
For this, we improve the closed-form expressions by Brand et al.~\cite{brand2020efficient} to get faster algorithm runtimes and extend the expressions to repeater schemes which involve distillation and cut-offs.
The runtime of the algorithm which evaluates these expressions is polynomial in the pre-specified size of the computed probability distribution's support.

In the second part of the paper, we optimize the choices of the cut-off to maximize the secret-key rate.
We study different cut-off strategies and find that the use of the optimal cut-off extends the parameter regime for which secret key can be generated and moreover significantly increases the secret-key rate for a large range of parameters.
We also analyze the dependence of the optimal cut-off on different properties of the hardware and find that memory quality highly influences the effectiveness of the cut-off, whereas the influence is small for success probability of entanglement swapping.
In addition, our numerical simulations show that for symmetric repeater protocols with evenly spaced nodes, a nonuniform cut-off (different cut-off time in different parts of the repeater chain) does not yield a significant improvement in end-to-end node secret key rate compared to a uniform cut-off.

This paper is organized as follows.
In section~\ref{sec:preliminaries}, we describe the class of repeater schemes under study and elaborate on the hardware model used in our simulations.
Section~\ref{sec:algorithms} presents the closed-form expressions and their evaluation algorithms for the waiting time distribution and output quantum states of repeater schemes which include cut-offs.
The second part of the work, on optimization of the cut-off, consists of section~\ref{sec:optimization}, where we provide details on the optimization procedure, and the results of the numerical optimization as presented in section~\ref{sec:results}.
Section~\ref{sec:discussion} ends our work with a conclusion.

\section{Preliminaries}
\label{sec:preliminaries}
\subsection{Class of repeater protocols considered}
\label{sec:protocol}
A quantum repeater chain connects two endpoints via several repeaters and aims to generate entanglement between the endpoints.
In this section, we elaborate on the class of quantum repeater chain protocols we study in this work, which is an extension of the class studied in~\cite{brand2020efficient} with the addition of cut-offs.
While doing so, we refer to both the endpoints and the repeater stations as nodes and to an entangled state between two nodes as a link.

\begin{figure}
    \includegraphics[width=0.98\linewidth]{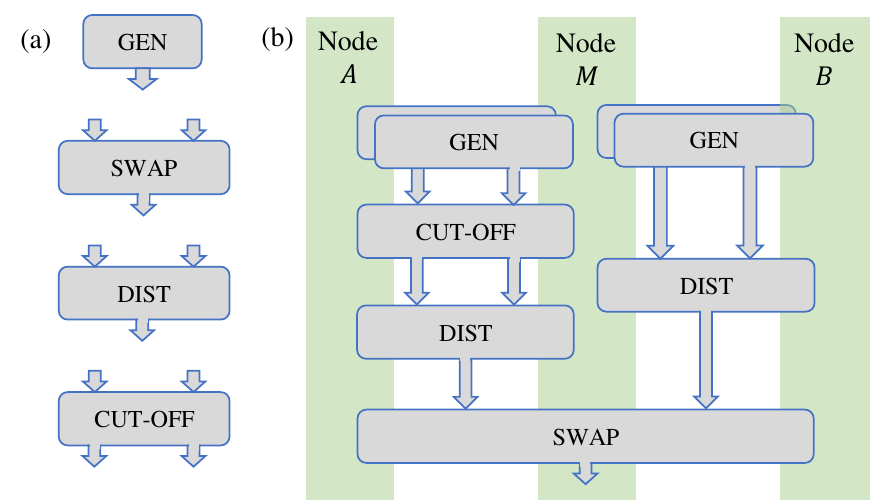}
    \caption{
    The class of repeater chain protocols considered in this work are composed of four different types of {\PUNIT}s.
    {\bf (a)} The four {\PUNIT}s: elementary-link generation between adjacent nodes (\GEN), entanglement swapping for connecting two short-distance links in a single long-distance one (\SWAP), entanglement distillation for converting two low-quality links in a single high-quality link (\DIST) and discarding two links (\CUTOFF), for example if their generation times differ by more than a pre-specified cut-off time.
    The repeater chain protocols we consider in this work are composed of combinations of the four {\PUNIT}s, provided that each {\CUTOFF} is succeeded by a {\SWAP} or {\DIST}.
    The in-/outgoing arrows of each {\PUNIT} indicate the number of entangled links that the block consumes/produces.
    {\bf (b)} An example of a composite protocol on three nodes (end nodes $A$ and $B$ and single repeater $M$).
    At the start of the protocol, two fresh elementary links are generated ({\GEN}) in parallel between adjacent nodes $A$ and $M$ and subsequently selected through a {\CUTOFF} block.
    The first two links that survive the cut-off are then distilled ({\DIST}) into a single link of higher quality.
    Asynchronously, the nodes $M$ and $B$ generate ({\GEN}) pairs of links until the distillation ({\DIST}) succeeds.
    Once distillation on both sides of node $M$ has succeeded, the resulting links $A\leftrightarrow M$ and $M\leftrightarrow B$ are converted via a {\SWAP} into a single entangled link between the end nodes $A$ and $B$.
    \label{fig:composite_protocol}
    }
\end{figure}

The class of quantum repeater protocols studied in this work are composed of the following four building blocks or {\PUNIT}s: elementary link generation (\GEN), entanglement swap (\SWAP), entanglement distillation (\DIST) and cut-off (\CUTOFF).
See \cref{fig:composite_protocol}(a).
All of these processes can fail, but the involved nodes receive a success or failure message.
That is, they are heralded.
In what follows, we describe these four {\PUNIT}s in more detail and subsequently explain how they can be composed into a repeater scheme that spans multiple nodes.

The first block {\GEN} represents the generation of fresh entanglement between two adjacent nodes.
We refer to those entangled pairs as an elementary link.
The {\GEN} block thus spans precisely two nodes, takes no input and outputs a single link.

The second and third blocks are entanglement swap ({\SWAP}) and entanglement distillation ({\DIST}).
In the setting of two nodes $A$ and $B$ with a middle station $M$ in between, an entanglement swap~\cite{zukowski1993} takes two links $A\leftrightarrow M$ and $M\leftrightarrow B$ and outputs a single link $A\leftrightarrow B$.
It spans at least three nodes.
Next, entanglement distillation probabilistically transforms two low-quality links between the same pair of nodes to a new one with higher quality \cite{bennett1996purification, deutsch1996quantum}.
The {\DIST} block thus spans at least two nodes, takes two links as input and outputs a single link, where each link is shared by the same pair of nodes.
Both {\SWAP} and {\DIST} consist of local operations including measurements and classical communication.
They can succeed or fail and in case of failure, both input links are lost.

The last {\PUNIT} is {\CUTOFF}, which takes two links as input (not necessarily between the same nodes).
It accepts or rejects the two input links depending on a success condition.
In case of success, it leaves the two input links untouched and outputs them again.
In case of failure, both input links are discarded.
In this work, we study three different success conditions.
In the first two, {\cutoffdifference} and {\cutoffmaximum}, `success' is declared if respectively the difference or the maximum of the input links' production times does not exceed some prespecified cut-off threshold.
In the third strategy, {\cutofffidelity}, the input states are passed on only if they are both of sufficient quality.
This success condition translates to a cut-off on the individual input states' fidelity with a maximally-entangled state (see \cref{sec:modelling}).

\begin{figure*}[t]
    \centering
    \includegraphics[width=\linewidth]{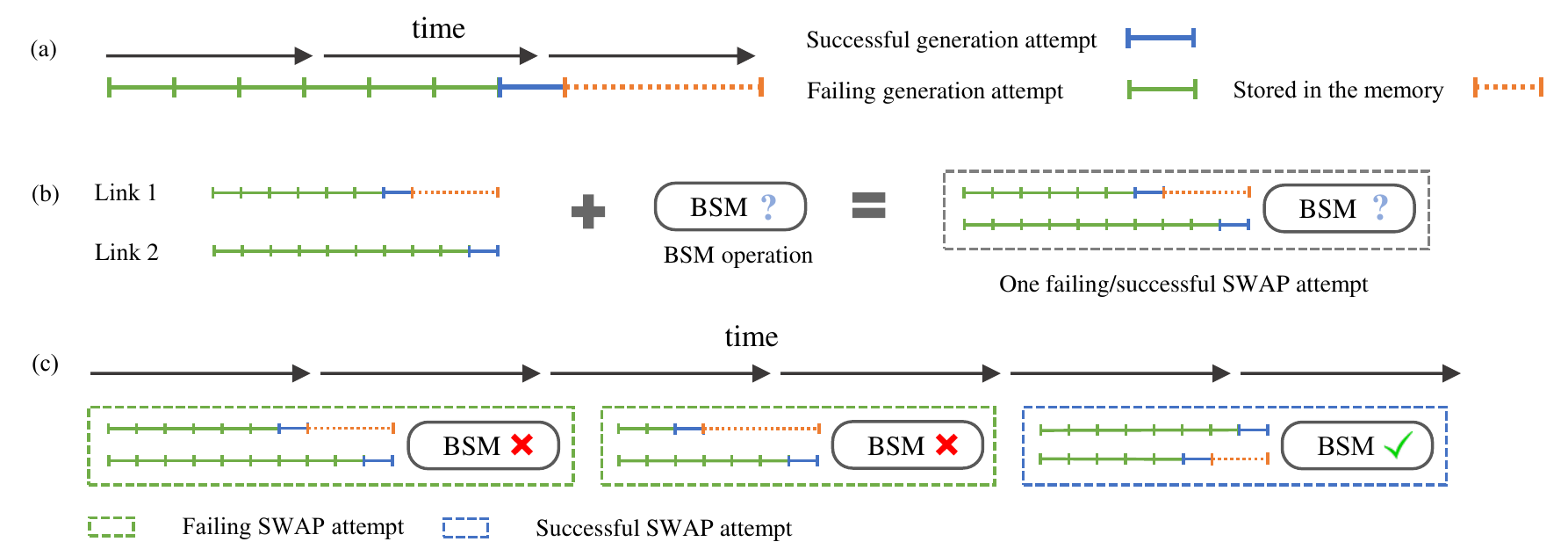}
    \caption{
Visualization of the waiting time until end-to-end entanglement is delivered for a 3-node repeater chain.
The repeater scheme consists of the generation of two elementary links, followed by an entanglement swap on the two links.
    {\bf (a)} A single link is generated in fixed-duration attempts, which succeed probabilistically and thus may fail (green line segment), after which generation is re-attempted until success (blue line segment).
    After that, the link is stored until it is consumed (dotted orange line segment).
    {\bf (b)} A run of the 3-node protocol until the first swap attempt, which consists of first preparing two input links in parallel, followed by a Bell state measurement (BSM).
    The link that is generated earlier than the other needs to wait in the memory (link 1 in the figure, the `waiting' is indicated by the dotted orange line).
While waiting, the earlier link's quality decreases due to decoherence.
    The total waiting time before the BSM equals the maximum of the generation times of the two links.
    The BSM operation can fail, in which case the two links are lost and need to be regenerated.
    {\bf (c)} A full run of the 3-node protocol, consisting of failed entanglement swaps (green dashed box) on fresh links until the first successful swap (blue dashed box).
    The total waiting time is the sum of the waiting times for the parallel generation of each pair of elementary links, up to and including the first successful swap.
    }
    \label{fig:attempt_until_success}
\end{figure*}

We now explain how the {\PUNIT}s described above can be composed into a single repeater protocol spanning a chain of nodes.
See \cref{fig:composite_protocol}(b) for an example.
Each composite protocol on a chain of nodes starts with one or multiple {\GEN} blocks between each pair of adjacent nodes for fresh elementary link generation.
A protocol then consists of stacking instances of the other three {\PUNIT}s in such a way that the output link(s) of one are used as input link(s) to the other.
The only restriction on how the {\PUNIT}s can be stacked is that both output links of {\CUTOFF} are used as inputs for one {\DIST} or {\SWAP} block.
As a consequence of the stacking, any repeater protocol in the class we study has a tree structure (see also \cref{fig:composite_protocol}b).
If a block at the root of a tree fails, then its input links are discarded and the {\GEN} blocks at the tree's leaves will restart.

We note that the class of repeater protocols described above includes, for instance, the well-known family of repeater schemes described by Briegel et al.~\cite{briegel1998quantum, briegel1999quantumrepeaters}.

\subsection{Model}
\label{sec:modelling}
We here describe how we model each of the four {\PUNIT}s described in \cref{sec:protocol}, which is identical to the modelling in~\cite{brand2020efficient}, except for the newly introduced {\CUTOFF} unit.
For each {\PUNIT}, we describe the success condition as well as the quantum state that it outputs.

First, we model the fresh entanglement generation ({\GEN}) using schemes which generate links in heralded attempts of duration $L_{\textnormal{internode}} / c$, where $L_{\textnormal{internode}}$ is the internode distance and $c$ is the speed of light in the used transmission medium, e.g.~ glass fiber~\cite{munro2015inside}.
We assume that each attempt is independent and succeeds with constant probability $0 < \pgen \leq 1$.
For simplicity, we assume that the nodes are equally spaced with internode distance $L_0$, so that each attempt in elementary link generation takes duration $\Delta t_0=L_0/c$, which will be the time unit in our numerical simulation.

We model the elementary link as a Werner state $\rho(w)$ with constant Werner parameter $w=w_0$ \cite{werner1989}:
\begin{equation}
    \rho(w) = w \dyad{\Phi^+}
    + (1-w)\frac{\mathbbm{1}_4}{4}
\label{eq:werner-state}
\end{equation}
where the Bell state
\begin{equation}
\ket{\Phi^+} = (\ket{00}+\ket{11})/\sqrt{2}
\label{eq:target-state}
\end{equation} is a maximally-entangled two-qubit state and 
\[
\mathbbm{1}_4 / 4 =
\left(\dyad{0} + \dyad{1}\right)
\otimes
\left(\dyad{0} + \dyad{1}\right) / 4
\] is the maximally-mixed state on two qubits.
We refer to the parameter $w$ with $0\leq w\leq 1$ as the Werner parameter.
Since a Werner state is completely determined by its Werner parameter, we use the Werner parameter to indicate the quantum state.

Equivalently to the Werner parameter, we will also express the state's quality using the fidelity, which for general density matrices $\rho$ and $\sigma$ is defined as
\[
    F(\rho, \sigma) := \Tr\left(\sqrt{\sqrt{\rho}\sigma\sqrt{\rho}}\right)^2.
    \]
The fidelity between a Werner state $\rho(w)$ and $\dyad{\Phi^+}$ equals
\begin{equation*}
    F = \frac{1+3w}{4}.
\end{equation*}

For the other {\PUNIT}s, the success conditions are summarized in \cref{tab:functions}.
In short: we model entanglement swapping ({\SWAP}) as succeeding with a constant probability $\pswap$.
For entanglement distillation ({\DIST}), we use the BBPSSW protocol \cite{bennett1996purification} which we adapt by bringing the output state back into Werner form.
The latter operation does not change the output state's fidelity with the target state $\ket{\Phi^+}$.
The success probability $\pdist$ of distillation is a function of the input states' Werner parameters (see \cite{brand2020efficient} for details).
The cut-off ({\CUTOFF}) success condition depends deterministically on the waiting time or the fidelity of the input links.

The states that any {\PUNIT} outputs are Werner states at any time of the execution of the protocol.
Indeed, a successful entanglement swap or distillation attempt maps Werner states to Werner states (see \cite{brand2020efficient} for a brief explanation).
Also, the {\CUTOFF} leaves the input states untouched in case of success, thereby outputting Werner states if it got those as input.
For each {\PUNIT}, the Werner parameters of the output links $\wout$ are a function of those of the input links and are given in \cref{tab:functions}.

In addition to the fact that the {\PUNIT}s change the quantum states they handle, the local quantum memories that are used to store the links are imperfect.
In our model, a link with initial Werner parameter $w$, which lives in memory for time $\Delta t$ until it is retrieved, decoheres to Werner parameter
\begin{equation}
    \label{eq:decay}
    w_{\textnormal{decayed}} = w \cdot e^{-\Delta t / \tcoh}.
\end{equation}
where $\tcoh$ is the joint coherence time of the two involved memories.

For simplicity, we ignore the time needed for classical communication between the nodes in this work as well as the time to perform the local operations.
The algorithm we provide can be easily extended to include these features, following the extension described in~\cite{brand2020efficient}.

In summary, for a given composite protocol (including the cut-off condition $\tau$ or $w_{\textnormal{cut}}$ for each {\CUTOFF} block), the simulation of the entanglement distribution process is determined by 4 hardware parameters: the success probability of elementary link generation $\pgen$, the swap success probability $\pswap$, the Werner parameter of the elementary link $w_0$ and the memory coherence time $\tcoh$.

\def\wAprime{w'_{A}}
\def\wBprime{w'_{B}}
\begin{table*}[t]
    \centering
    \caption{Overview of success probability and the output Werner parameter for each {\PUNIT}}
    \label{tab:functions}
    \begin{tabular}{l ll}
        \toprule
        {\PUNIT} & success probability $p$  & Werner parameter $\wout$
    \\ \midrule
        generation (\GEN) & $\pgen$ (constant)  & $w_0$
    \\ [1.3ex]
        entanglement swapping (\SWAP) & $\pswap$ (constant)& $ \wAprime \cdot \wBprime $
    \\ [1.3ex]
        entanglement distillation (\DIST) & 
        $\displaystyle  \pdist = \frac{1 + \wAprime\wBprime}{2} $ & 
    
        $\displaystyle  \frac{\wAprime + \wBprime + 4\wAprime\wBprime}{6\pdist}$
    \\
        \cutoffdifference & 
        $\left.\pto=\begin{cases}
            1 & \textnormal{ if }  |\tA-\tB| \le \tau \\ 
            0 & \textnormal{ otherwise}
        \end{cases}
        \right.
        $
        &
        $\wAprime, \; \wBprime$
        \vspace{3mm}
    \\

        \cutofffidelity & 
        $\left.\pto=\begin{cases}
            1 & \textnormal{ if } \wAprime \ge w_{\textnormal{cut}}
                \textnormal{ and } \wBprime \ge w_{\textnormal{cut}} \\ 
            0 & \textnormal{ otherwise}
        \end{cases}
        \right.
        $
        &
        $\wAprime, \; \wBprime$
    \\ [3.7ex]
        \vspace{3mm}
        \cutoffmaximum & 
        $\left.\pto=\begin{cases}
            1 & \textnormal{ if} \max(t_A, t_B) \le \tau \\ 
            0 & \textnormal{ otherwise}
        \end{cases}
        \right.
        $
        &
        $\wAprime, \; \wBprime$
    \\
    \midrule
    \multicolumn{3}{l}{
    \begin{minipage}{0.7\textwidth}
    where $(\tA, \wA)$ and $(\tB, \wB)$ are the waiting time and Werner parameter of the links $A$ and $B$ provided as input to the {\PUNIT}.
    Parameters $\tau$ and $w_{\textnormal{cut}}$ are the cut-off thresholds on time and Werner parameter, respectively.
    The primed notation denotes Werner parameter with decay in \cref{eq:decay} applied to the link that waits until the other is finished: $w'_X = w_X\cdot e^{-|\tA - \tB|/\tcoh}$ if $t_X = \min(\tA, \tB)$ and $w'_X=w_X$ otherwise, for $X \in \{A, B\}$.
        For an explanation of the different {\PUNIT}s, see \cref{sec:protocol}.
    \end{minipage}
    }
    \\
        \\ [-2.5ex]
        \bottomrule
    \end{tabular}
\end{table*}
\

\subsection{Waiting time and produced end-to-end state in repeater schemes using probabilistic components}
\label{sec:metrics}
In this work, we study the time until the first entangled pair of qubits is generated between the end nodes of the repeater chain (called `waiting time' from here on) and the state's quality, expressed as its Werner parameter (recall that the end-to-end state is a Werner state, see~\cref{sec:modelling}).
Because the repeater chain protocols we study in this work are composed of probabilistic components, both the waiting time and the end-to-end state's Werner parameter are random variables.
For an illustration of the random behavior of the waiting time, see \cref{fig:attempt_until_success}.
We characterize the quality by the averaged Werner parameters of all states generated at the same time step $t$.
The algorithm we present in this work computes the probability distribution $\Pr(T=t)$ of the waiting time $T$ and the average Werner parameter $W(t)$ of the end-to-end state which is delivered at time $t$.

We finish this section by noting that by considering the average Werner parameter, we ignore the `history' of a link, resulting in a suboptimal estimation of the fidelity of the states.
To see this, consider for example the three-node protocol of \cref{fig:composite_protocol}(b).
In this protocol, the following two series of events lead to an output entangled pair between nodes $A$ and $B$ at time $t=10$: (i) all {\GEN} blocks fail at each timestep $t<10$ but succeed at time $t=10$, after which all other {\PUNIT}s also succeed immediately, (ii) the {\PUNIT}s between $A$ and $M$ all succeed at time $t=1$, while the {\GEN} blocks between $M$ and $B$ succeed at time $t=10$, followed by all other remaining {\PUNIT}s also succeeding at time $t=10$.
In case (i), no entanglement has waited in memory, whereas in case (ii), the produced link between $A$ and $M$ has waited $10$ timesteps and decohered in that time.
By keeping track of the timestamps at which the several {\PUNIT}s succeeded, one could distinguish these two scenarios.
Since the resulting fidelity estimation computation is rather complex and in this work, we focus on quantifying the effect of a cut-off, we leave such advanced fidelity estimation for future work.

\section{Computing the waiting time distribution and the output Werner parameter}
\label{sec:algorithms}

\begin{figure}[t]
    \centering
    \includegraphics[width=\linewidth]{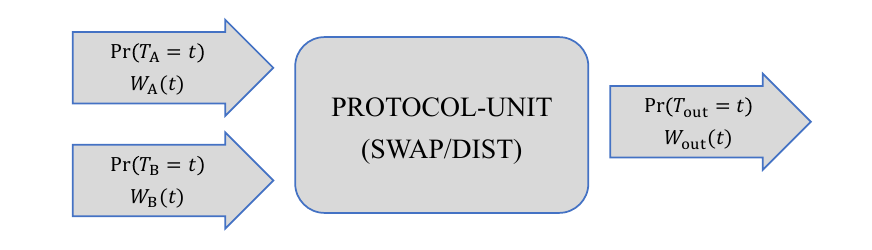}
    \caption{The workflow of the algorithm for one {\PUNIT} ({\SWAP} or {\DIST}).
    It takes the waiting time distribution and Werner parameter of the two input links and computes those of the output.
}
    \label{fig:algorithm}
\end{figure}

In this section, we present closed-form expressions of the waiting time probability distribution and Werner parameter of the output links for each {\PUNIT}, as function of waiting time distribution and Werner parameter of its input links.
Expressions for a composite protocol are obtained by iterative application over the {\PUNIT}s that the protocol consists of.
These expressions naturally lead to an algorithm for their evaluation, which we also present in this section.

Closed-form expressions for {\GEN} and {\SWAP} were already obtained by Brand {\it et al.}~\cite{brand2020efficient}, who explicitly mentioned that their approach does not generalize straightforwardly to {\DIST}.
Here, we include {\DIST} and even {\CUTOFF}, provided the latter is succeeded by {\SWAP} or {\DIST}.
The novel idea is to use separate expressions for the waiting time probability distribution of a successful and failed attempt.
We then express the total waiting time distribution and the Werner parameter as those of the successful attempt averaged by the occurrence probability of all possible sequences of failed attempts, where the weighted average is efficiently computed using convolution.
As an additional benefit, the evaluation algorithm for {\SWAP} is faster than the one presented by Brand et al.

In the following, we first derive general closed-form expressions for the waiting time distribution and Werner parameter of one {\PUNIT} in \cref{sec:repe_attmpts}.
We then give specific expressions for each {\PUNIT} individually in \cref{sec:alg gen,sec:alg swap,sec:alg dist,sec:alg timeout}.
In the last section (\cref{sec:alg complexity}), we show how these expressions can be converted into an efficient algorithm.
We also explain how to modify the closed-form expressions using the discrete Fourier transform, motivated by its use in~\cite{kuzmin2019scalable, kuzmin2021diagrammatic}.
These modified expressions lead to an even faster algorithm for computing the waiting time and Werner parameter, which we provide in Appendix~\ref{sec:complexity improvement}.
We denote the random variables of the waiting time and average Werner parameter as $T$ and $W(t)$, with subscript A and B for the input links and `out' for the output link (see \cref{fig:algorithm}).

\subsection{General closed-form expressions for waiting time and produced states for all protocol-units}
\label{sec:repe_attmpts}
\subsubsection{Random variable expression for the waiting time of {\PUNIT}s}
We start by presenting an expression for the random variable $\Tout$.
To study the waiting time distribution, we divide the total waiting time into the waiting time for each attempt.
An attempt can fail or succeed and it repeats until the first successful attempt occurs (see \cref{fig:attempt_until_success}).
The total waiting time $\Tout$ is given by
\begin{equation}
    \label{eq:compound sum}
    \Tout = \sum_{i=1}^K M^{(i)}
\end{equation}
where $M^{(i)}$ are i.i.d.
random variables characterizing the waiting time of each attempt and therefore each is a function of the waiting time of two input links $\TA$, $\TB$.
For example, for {\SWAP}, we have $M=\max(\TA, \TB)$, {\it i.e.}~we need to wait until both links are ready to perform the operation.
$K$ is the number of attempts we need until the first successful attempt occurs, which is also a random variable.

The success or failure of one attempt is characterized by a probability $p$.
The success probability $p$ of one attempt is independent of that of others and is given by $p=p(\tA, \tB, \wA, \wB)$ (\cref{tab:functions}).
We reduce the Werner parameter dependence to time dependence by plugging in $w_A=\WA(t_A)$ and $w_B=\WB(\tB)$.
Hence, we write $p(\tA, \tB)$ in the rest of this section.

The time dependence of $p$ implies that, in general, $K$ is correlated to $M^{(j)}$.
To make this correlation between $K$ and $M$ in \cref{eq:compound sum} explicit, we introduce a random variable $Y$. $Y$ denotes the binary random variable describing success (1) or failure (0) of a single attempt, subjected to the success probability $p(\tA, \tB)$.
The time-dependent success probability can be understood as the success probability with given waiting time $\tA$, $\tB$ of the input links:
\begin{equation*}
    p(\tA, \tB) = \Pr(Y=1 | \TA=\tA, \TB=\tB).
\end{equation*}
We then rewrite \cref{eq:compound sum} with a sum over all possible number of attempts weighted by its occurrence probability\cite{collins2007multiplexed}:
\begin{equation}
    \label{eq:semi-goem random variable T}
    \Tout = \sum_{k=1}^\infty
    \left\{
        \left(
            Y^{(k)}\prod_{j=1}^{k-1}
            \left(
                1-Y^{(j)}
            \right)
        \right)
        \cdot
        \sum_{i=1}^{k}
        M^{(i)}
    \right\}
    .
\end{equation}
The expression in round brackets evaluates to 1 precisely if $Y^{(k)} = 1$ and $Y^{(j)} = 0$ for all $j < k$, and to 0 in all other cases.
This factor thus makes that only the sum $\sum_{i=1}^k M^{(k)}$ is taken for which $k$ is the first successful attempt.
Notice that $Y^{(j)}$ and $M^{(i)}$ are correlated for all $i=j$ because they describe the same attempts.
In the next section, we go further to compute the probability distribution of $\Tout$.

\subsubsection{A closed-form expression for the waiting time distribution}
In the following, we give an expression of the waiting time distribution $\Pr(\Tout=t)$ for one {\PUNIT}.

We consider the generation time of a successful or failed attempt separately and use the joint distribution of $M$ and $Y$. We define the joint distribution that one attempt succeeds/fails and takes time $t$ as
\begin{align}
    \label{eq:semi-geom Ps}
    \Ps(t)
    \coloneqq &
    \Pr(M=t, Y=1) \nonumber \\
    = &
    \smashoperator{
        \sum_{
            \quad \quad \quad \quad
            \tA, \tB: \max(\tA, \tB) = t
        }
    }
    \Pr(\TA=\tA,\TB=\tB) \cdot p (\tA, \tB),
\end{align}
\begin{align}
    \label{eq:semi-geom Pf}
    \Pf(t)
    \coloneqq &
    \Pr(M=t, Y=0) \nonumber \\
    = &
    \smashoperator{
        \sum_{
            \quad \quad \quad \quad
            {\tA, \tB: \max(\tA, \tB) = t}
        }
    }
    \Pr(\TA=\tA, \TB=\tB) \cdot 
    [1 - p](\tA, \tB)
    .
\end{align}
In the above equation, we iterate over all possible combinations of the input links' generation time $\tA$, $\tB$ that leads to a waiting time $t$ for this attempt.

With the definition \cref{eq:semi-geom Ps} and \cref{eq:semi-geom Pf}, the sum of the waiting time for all attempts can be obtained by
\begin{empheq}[box=\fbox]{equation}
    \label{eq:semi-geom iterative convolution T}
    \Pr(\Tout=t)=
    \sum_{k=1}^\infty
    \left[
        \left(
            \Conv\limits_{j=1}^{k-1} \Pf^{(j)} 
        \right)
        \conv \Ps
    \right]
(t)
\end{empheq}
where $\conv$ is the notation for convolution and the sum over $k$ considers all the possible numbers of attempts.
The notation $\Conv\limits_{j=1}^{k-1} \Pf^{(j)}$ represents the convolution of $k-1$ independent functions $\Pf$.
In the above equation, the discrete linear convolution is defined by
\begin{equation}
    \label{eq:convolution}
    [f_1\conv f_2](t) = \sum_{t'=0}^t f_1(t-t')\cdot f_2(t').
\end{equation}
If $f_1$, $f_2$ describe two probability distributions of two random variables, their convolution is the distribution of the sum of those two random variables.
However, neither $\Pf$ or $\Ps$ characterizes a random variable since they are joint distributions including $Y$.
That is to say, $\Ps$ and $\Pf$ do not sum up to 1.
Instead, we have
\begin{equation*}
    \sum_{t} \Pf(t) + \sum_{t} \Ps(t) = 1.
\end{equation*}
Therefore, the convolution here cannot be simply interpreted as a sum of two random variables.
Instead, it is the summed waiting time conditioned on the success/failure of each attempt.

As we will show in sec.~\ref{sec:alg complexity}, eq.~\cref{eq:semi-geom iterative convolution T} is sufficient for the derivation of the main algorithm for computing the probability distribution of $\Tout$ we present in this work.
The algorithm's runtime is partially determined by the sum and the convolution in the summand in eq.~\cref{eq:semi-geom iterative convolution T}.
Fortunately, these can be eliminated by the use of the discrete Fourier transform, resulting in a faster alternative algorithm.
Below, we use the Fourier transform to derive an equivalent expression to eq.~\cref{eq:semi-geom iterative convolution T}.
The alternative algorithm is given in Appendix~\ref{sec:complexity improvement}

Since the discrete Fourier transform acts on a finite sequence of numbers, we first truncate the probability distribution at a fixed time $L$, {\it i.e.}~we obtain the finite sequence $\{\Pr(\Tout = t) | t = 0, 1, 2, \dots, L\}$.
If $\vec{x}:=x_0, x_1, \dots, x_{L-1}$ is a sequence of complex numbers, then its Fourier transform $\mathcal{F}(\vec{x})$ is the sequence $y_0, y_1, \dots, y_{L-1}$ given by
\begin{equation}
    \label{eq:fourier-transform}
    y_j = \sum_{k=0}^{L-1} x_k \cdot \exp\left(-2\pi \mathrm{i}\cdot j \cdot k/L\right)
\end{equation}
where $\mathrm{i}$ is the complex unit.
The Fourier transform is a linear map and moreover it converts convolutions into element-wise multiplication, {\it i.e.}~$\mathcal{F}(\vec{x} \conv \vec{x'}) = \mathcal{F}(\vec{x}) \cdot \mathcal{F}(\vec{x}')$.
As a consequence, taking the Fourier transform of both sides of eq.~\cref{eq:semi-geom iterative convolution T} yields

\begin{equation*}
    \mathcal{F}\left[\Pr(\Tout=t)\right]=
    \sum_{k=1}^\infty
    \left[
        \left(
            \prod\limits_{j=1}^{k-1} \mathcal{F}(\Pf)^{(j)}
        \right)
        \cdot \mathcal{F}[\Ps]
    \right]
    (t)
.
\end{equation*}
Because, $\Pf^{(j)}$ are identical distribution for all $j$, we use the identity $\sum_{k=1}^{\infty}x^{(k-1)} = 1/(1-x)$ to obtain
\begin{equation}
    \label{eq:fourier T}
    \Pr(\Tout=t)=
    \mathcal{F}^{-1}
    \left[
    \frac{\mathcal{F}[\Ps]}{1 - \mathcal{F}[\Pf]}
    \right]
    (t)
    .
\end{equation}

\subsubsection{A closed-form expression for the Werner parameter}
Here, we derive the expression for the Werner parameter $\Wout(t)$.

\def\tsuc{t}
To arrive at $\Wout(t)$, we first compute the average Werner parameter of the output link of one attempt, given that it succeeds and finishes at time $\tsuc$:
\begin{align}
    \label{eq:semi-geom Wsuc}
    \Wprep(\tsuc)
    =
    \frac{
    \quad
    \sum \limits_{\mathclap{\quad \quad \quad \quad \quad \tA,\tB: \max(\tA,\tB) = \tsuc}}
    \Pr(\TA=\tA, \TB=\tB) \cdot
    [p \cdot \wout] (\tA,\tB)
    }
    {
        \Ps(\tsuc)
    }
.
\end{align}
Here, $\wout$ is the Werner parameter of the output link of a successful attempt and $p$ the success probability (\cref{tab:functions}).
We again simplify the notation with $\wout(\tA, \tB)=\wout(\tA, \tB, \WA(\tA), \WB(\tB))$.

Next, we take a weighted average of $\Wsuc$ over all possible sequences of failed attempts, followed by a single successful attempt:
\begin{empheq}[box=\fbox]{equation}
    \label{eq:semi-geom iterative convolution W}
    \Wout(t) =
    \frac{
        \sum\limits_{k=1}^\infty
        \left[
            \left(
                \Conv\limits_{j=1}^{k-1} \Pf
            \right)
            \conv
            \left(
                \Ps \cdot \Wprep
            \right)
        \right]
        (t)
    }
    {
        \Pr(\Tout=t)
    }.
\end{empheq}
where $\Conv\limits_{j=1}^{k-1} \Pf^{(j)}$ computes the waiting time distribution of $k-1$ failed attempts and the additional convolution is the weighted average.

For eq.~\eqref{eq:semi-geom iterative convolution T}, which is an expression for the probability distribution of $\Tout$, we obtained a more compact equivalent, eq.~\eqref{eq:fourier T}, by moving to Fourier space.
By an analogous derivation, we can get a more compact expression for $\Wout$ than eq.~\eqref{eq:semi-geom iterative convolution W}:
\begin{equation}
    \label{eq:fourier W}
    \Wout(t) =
    \mathcal{F}^{-1}
    \left[
    \frac
    {
        \mathcal{F}[\Ps \cdot \Wprep]
    }
    {
        1-\mathcal{F}[\Pf]
    }
    \right]
    \frac{1}
    {
        \Pr(\Tout=t)
    }
    (t)
    .
\end{equation}

\subsection{Specific case: GEN}
\label{sec:alg gen}
We give here the expression for {\PUNIT} {\GEN}. Since {\GEN} does not have input links, the output does not rely on the expression introduced in the \cref{sec:repe_attmpts}. Because one attempt in {\GEN} takes one time step and the success probability $\pgen$ is a constant, the waiting time can be described by a geometric distribution
\begin{equation*}
    \Pr(\Tout=t) = \pgen (1-\pgen)^{t-1}.
\end{equation*}
The output state is a Werner state with Werner parameter $w_0$ as described in \cref{sec:modelling}.

\subsection{Specific case: SWAP}
\label{sec:alg swap}

For entanglement swap, since $\pswap$ is constant, $Y$ is not correlated with $M$. As a result, $\Ps$ and $\Pf$ differ only by a constant coefficient (see \cref{eq:semi-geom Ps} and \cref{eq:semi-geom Pf}).
Therefore, we can factor the constant out and get
\begin{align*}
    \Pr(\Tout=t)
    &=
    \sum_{k=1}^{\infty}
    \pswap(1-\pswap)^{k-1}
    \left[
        \Conv\limits_{j=1}^k m
    \right]
\end{align*}
where
\begin{equation*}
m(t) \coloneqq Pr(M=t) = 
\smashoperator{
    \sum_{
        \quad \quad \quad \quad
        \tA, \tB: \max(\tA, \tB) = t
    }
}
\Pr(\TA=\tA,\TB=\tB).
\end{equation*}
This is exactly the geometric compound distribution obtained in \cite{brand2020efficient}.

For the Werner parameter, we can directly use \cref{eq:semi-geom iterative convolution W} and obtain
\begin{equation}
    \label{eq:entanglement swap W}
    \Wout=
    \sum_{k=1}^{\infty}
    \pswap(1-\pswap)^{k-1}
    \left[
        \left(
        \Conv\limits_{j=1}^{k-1} m 
        \right)
        \conv 
        \left(
            m \cdot \Wsuc
        \right)
    \right]
    .
\end{equation}
Compared to the expression in \cite{brand2020efficient}, this expression replaces the iteration over all pair of possible input Werner parameters for each $k$ by convolution.

Both expressions above can also be written in Fourier space by substituting $\Ps=\pswap m(t) $ and $\Pf=(1-\pswap) m(t) $ in \cref{eq:fourier T,eq:fourier W}.

\subsection{Specific case: DIST}
\label{sec:alg dist}
For entanglement distillation, the success probability depends on the Werner parameters.
As discussed in \cref{sec:repe_attmpts}, we can compute $\Tout$ and $\Wout$ because we iterate over all possible combinations of $\tA$ and $\tB$ and we use $W(t)$ to reduce the dependence on Werner parameters to the dependence on the waiting time.
The calculation goes as follows.
First, we compute $\Pf$ and $\Ps$ using $p(t_A, t_B) = \pdist(W(t_A), W(t_B))$ (\cref{tab:functions}).
Then, we plug in $\Pf$ and $\Ps$ in \cref{eq:semi-geom iterative convolution T} to compute $\Tout$.
Finally, $\Wout$ can be calculated similarly using \cref{tab:functions}, \cref{eq:semi-geom Wsuc,eq:semi-geom iterative convolution W}.

\subsection{Specific case: CUT-OFF}
\label{sec:alg timeout}
{\CUTOFF} selects the input links and accepts them if the cut-off condition described in \cref{sec:modelling} is fulfilled.
We consider only the case where {\CUTOFF} is followed by {\SWAP} or {\DIST}, so that the two blocks together output a single entangled link.

\subsubsection{The waiting time distribution}
We define a new binary variable $\Yt$ representing whether the cut-off condition is fulfilled.
The corresponding success probability is described by $\pto$ in \cref{tab:functions}.
In addition, we also define the waiting time of one cut-off attempt as $Z$, in contrast to $M$ for a swap or distillation attempt.
For {\CUTOFF}, we need to distinguish the waiting time of a successful and a failed attempt.
In the case of success, we always have $\Zs=\max(\TA,\TB)$, {\it i.e.}~we wait until two links are produced.
However, in the case of failure, the waiting time is different for different cut-off strategies.
With the notation $\Zf=t_{\textnormal{fail}}(\TA,\TB)$, we have the following:
for \cutoffdifference, $t_{\textnormal{fail}}(\TA,\TB)=\min(\TA,\TB) + \tau$, because there is no need to wait for the second link longer than the cut-off threshold.
For \cutoffmaximum, $t_{\textnormal{fail}}(\TA,\TB)$ is the constant $\tau$, {\it i.e.}~the maximal allowed waiting time.
For \cutofffidelity, it is $t_{\textnormal{fail}}(\TA,\TB)=\max(\TA,\TB)$.

Similar as the nested structure shown in \cref{fig:attempt_until_success}, a swap or distillation attempt is now composed of several cut-off attempts. We can write its waiting time $M$ as
\begin{equation*}
    M = \sum_k
    \left\{
        \left
            [\Yt^{(k)}\prod_{j=1}^{k-1}
            \left(
                1-\Yt^{(j)}
            \right)
        \right]
        \cdot
        \left[\Zs^{(k)} + \sum_{i=1}^{k-1} 
            \left(
                \Zf^{(i)}
            \right)
        \right]
    \right\}
\end{equation*}
This expression will replace $M=\max(\TA, \TB)$ used in \cref{eq:semi-goem random variable T}.
For $\tau=\infty$ or $w_{\textnormal{cut}}=0$, {\it i.e.}~no cut-off, $\Yt$ is always 1.
Therefore, $k=1$ is the only surviving term and the two expressions coincide.

To calculate the waiting time distribution, we need three joint distributions: $\Pf'$ for unsuccessful input link preparation because of the cut-off, $P'_{\textnormal{s,f}}$ for successful preparation but unsuccessful swap/distillation and $P'_{\textnormal{s,s}}$ for both successful:
\begin{align*}
    \Pf'(t)= 
    & \Pr(M=t, \Yt=0) \nonumber \\
    =
    & \smashoperator{
        \sum_{
            \quad \quad \quad \quad \quad
            \tA, \tB: t_{\textnormal{fail}}(\tA, \tB) = t
            }
    }
    \Pr(\TA=\tA, \TB=\tB) \cdot
    [1-\pto](\TA,\TB)
\end{align*}
\begin{align*}
    P'_{\textnormal{s,f}}(t)= 
    & \Pr(M=t, \Yt=1, Y=0) \nonumber \\
    =
    &\smashoperator{
        \sum_{
            \quad \quad \quad \quad
            {\tA, \tB: \max(\tA,\tB) = t}
            }
    }
    \Pr(\TA=\tA, \TB=\tB) \cdot
    [
        \pto \cdot
        (1-p)
    ]
    (\tA,\tB)
\end{align*}
\begin{align*}
    P'_{\textnormal{s,s}}(t)= 
    & \Pr(M=t, \Yt=1, Y=1) \nonumber \\
    =
    &\smashoperator{
        \sum_{
            \quad \quad \quad \quad
            {\tA, \tB: \max(\tA,\tB) = t}
            }
    }
    \Pr(\TA=\tA, \TB=\tB) \cdot
    [
        \pto \cdot
        p
    ]
    (\tA,\tB)
    .
\end{align*}
The prime notation indicates that they describe the waiting time of one attempt in {\CUTOFF}, in contrast to one attempt in swap or distillation.

For one attempt in swap/distillation with time-out, we then get similarly to \cref{eq:semi-geom iterative convolution T} 
\begin{align*}
    \Ps(t) = \Pr(M=t,Y=1) =
    \sum_k
    \left[
        \left(
            \Conv\limits_{j=1}^{k-1} \Pf'^{(j)}
        \right)
        \conv P'_{\rm{s,s}}
    \right]
    (t) \\
    \Pf(t) = \Pr(M=t,Y=0) =
    \sum_k
    \left[
        \left(
            \Conv\limits_{j=1}^{k-1} \Pf'^{(j)}
        \right)
        \conv P'_{\rm{s,f}}
    \right]
    (t)
\end{align*}
as well as the expressions in Fourier space analogous to \cref{eq:fourier T}
\begin{align*}
    \Ps(t) = \Pr(M=t,Y=1) =
    \mathcal{F}^{-1}
    \left[
    \frac{\mathcal{F}[P'_{\textnormal{s,s}}]}{1 - \mathcal{F}[\Pf']}
    \right]
    ,
    \\
    \Pf(t) = \Pr(M=t,Y=0) =
    \mathcal{F}^{-1}
    \left[
    \frac{\mathcal{F}[P'_{\textnormal{s,f}}]}{1 - \mathcal{F}[\Pf']}
    \right]
    .
\end{align*}
The total waiting time then follows by substituting the expressions for $\Pf$ and $\Ps$ above in \cref{eq:semi-geom iterative convolution T} or \cref{eq:fourier T}.

For entanglement swap, {\it i.e.}~constant success probability $\pswap$, simplification can be made for this calculation.
In this special case, $P'_{\textnormal{s,f}}$ and $P'_{\textnormal{s,s}}$ differ only by a constant and the same holds for $\Ps$ and $\Pf$.

\subsubsection{The Werner parameter}
\label{sec:cutoff werner}
For the Werner parameter, we now need three steps.

We start from calculating the resulting Werner parameter of a swap or distillation for the very last preparation attempt where $\Yt=Y=1$.
It is denoted by $\Wsuc$ and we only need to replace $\Ps$ by $P'_{\textnormal{s,s}}$ and $p \cdot \wout$ by $\pto \cdot p \cdot \wout$ in \cref{eq:semi-geom Wsuc}.

Next, we compute the Werner parameter $\Wprep(t)$ as a function of time $t$ that includes the failed cut-off attempts, in analog to the derivation of eq.~\cref{eq:semi-geom iterative convolution W}.
$\Wprep(t)$ is the Werner parameter that the pair of output links of {\CUTOFF} will produce, given that the swap or distillation operation following is successful:
\begin{align*}
    \Wprep(t)
    &=
    \frac
    {
        \sum\limits_{k=1}^\infty
        \left[
            \left(
                \Conv\limits_{j=1}^{k-1} \Pf'
            \right)
            \conv (P'_{\rm{s,s}} \cdot \Wsuc)
        \right]
        (t)
    }
    {
        \Ps(t)
    }.
\end{align*}
Finally, we consider the time consumed by failed attempts in {\SWAP} or {\DIST} and obtain
\begin{align*}
    \Wout(t)
    &=
    \frac
    {
        \sum\limits_{k=1}^\infty
        \left[
            \left(
                \Conv\limits_{j=1}^{k-1} \Pf
            \right)
            \conv (\Ps\cdot \Wprep)
        \right]
        (t)
    }
    {
        \Pr(\Tout=t)
    }
    .
\end{align*}
Using the Fourier transform, the two expressions above become
\begin{align*}
    \Wprep(t)
    &=
    \mathcal{F}^{-1}
    \left[
    \frac
    {
        \mathcal{F}[P'_{\textnormal{s,s}} \cdot \Wsuc]
    }
    {
        1 - \mathcal{F}[\Pf']
    }
    \right]
    \frac{1}
    {
        \Ps
    },
\\
    \Wout(t)
    &=
    \mathcal{F}^{-1}
    \left[
    \frac
    {
        \mathcal{F}[\Ps \cdot \Wprep]
    }
    {
        1 - \mathcal{F}[\Pf]
    }
    \right]
    \frac{1}
    {
        \Pr(\Tout=t)
    }
    .
\end{align*}

\subsection{Converting the closed-form expressions into an efficient algorithm}
\label{sec:alg complexity}
In the sections above, we presented closed-form expressions for $\Tout$ and $\Wout$ for each of the four {\PUNIT}s, as a function of waiting time distribution and Werner parameter of the input links.
In order to convert these expressions into an algorithm, we take the same approach as in~\cite{brand2020efficient} and cap the infinite sum in \cref{eq:semi-geom iterative convolution T,eq:semi-geom iterative convolution W} by a pre-specified truncation time $\ttr$.
This yields a correct $\Pr(\Tout = t)$ and $\Wout(t)$ for $t\in \{1, \dots, \ttr\}$ since in each of the expressions with an infinite sum above, $\Pr(\Tout = t)$ and $\Wout(t)$ are only dependent on waiting time and Werner parameter of input links produced at time $t' \leq t$.

We now show that the algorithm scales polynomially in terms of $\ttr$.
To analyze the complexity, we divide the algorithm into two parts: computing the distribution for one attempt, {\it i.e.}~the iteration over all possible values of $\TA$, $\TB$ (\cref{eq:semi-geom Pf,eq:semi-geom Ps,eq:semi-geom Wsuc}) and for the whole \PUNIT (\cref{eq:semi-geom iterative convolution T,eq:semi-geom iterative convolution W}).

The complexity for the first part is $\mathcal{O}(\ttr^2)$ since it iterates over two discrete random variables up to $\ttr$.
For the second part, because we need at least one time step in each attempt, {\it i.e.}~$\Pr(T=0)=0$, only the first $\ttr$ convolutions will have non-zero contribution.
We can perform the convolution iteratively for each $k$ using at most $\ttr$ convolutions.
The complexity of one convolution with fast Fourier transform (FFT) is $\mathcal{O}(\ttr\log{\ttr})$ \cite{cooley1965algorithm}.
Thus, the complexity of the second part scales as $\mathcal{O}(\ttr^2\log{\ttr})$.
The overall complexity, therefore, is $\mathcal{O}(\ttr^2\log{\ttr})$.

In \cref{sec:complexity improvement}, we show that with further simplification of \cref{eq:semi-geom Pf,eq:semi-geom Ps} as well as expressions in Fourier space (equations \cref{eq:fourier T,eq:fourier W}), the complexity can be reduced to $\mathcal{O}(\ttr\log{\ttr})$, with an exponentially vanishing error.

The preceding discussion shows that the algorithm is efficient as a function of the truncation time.
However, for fixed truncation time, the probability mass captured by the algorithm decreases as the number of nodes increases.
For protocols without cut-off, variations of the arguments in~\cite{brand2020efficient} would allow to prove that the algorithm introduced here is also efficient for fixed probability mass.
Unfortunately, the arguments do not translate to protocols with cut-off.
This is because for these protocols, the truncation time that covers a fixed probability mass can grow exponentially with the number of nodes, {\it i.e.}~such an algorithm can not exist.

As an example, consider a nested protocol on $2^n$ repeater segments ($n=0,1,2, \dots$), which for $n=1$ consists of a {\GEN} block only, and for each additional level $n>1$, each pair of adjacent links is connected by a {\CUTOFF} followed by a {\SWAP}.
We set $\tau = 0$ for each cut-off, {\it i.e.}~all elementary links need to be generated at the same time and also all entanglement swaps should succeed at the first attempt for the links to survive all the cut-offs.
Since $2^{n}$ elementary links need to be generated and the protocol consists of $2^n - 1$ swaps, the probability of successful end-to-end entanglement before time $t$ equals $1 - (1 - p)^t$ with $p = \pgen ^ {N - 1} \cdot \pswap ^ {N - 2}$, {\it i.e.}~decreases exponentially in the number of nodes $N = 2^n + 1$.

\section{Optimization}
\label{sec:optimization}
In this section, we describe the details of our optimization over cut-offs, including the figure of merit and optimization method.

In our numerical study, we use the secret-key rate of the BB84 protocol~\cite{bennett1984quantum} as a figure of merit to assess the performance of composite repeater protocols.
We compute the secret-key rate $R$ as the secret-key fraction divided by the average waiting time
\begin{equation}
    \label{eq:secret key rate}
    R = \frac{r}{\bar{T}}.
\end{equation}
The secret-key fraction $r$ describes the amount of secret key that can be extracted from the generated entanglement and is given by~\cite{shor2000simple, lo2005efficient}
\[
    r(w) = \max
    \left\{
        0, 1 - h[e_X(w)] - h[e_Z(w)]
    \right\}
\]
where $h(p) = -p \log_2(p) - (1 - p) \log_2 (1 - p)$ is the binary entropy function and $e_X$ ($e_Z$) is the quantum bit error rate in the $X$ ($Z$) basis.
Since the quantum states tracked by our algorithm are Werner states at any point in the execution of the composite repeater protocol (see \cref{sec:preliminaries}), the quantum bit error rate can be expressed as function of the end-to-end state's Werner parameter:
\[
    e_Z(w) = \bra{01} \rho(w) \ket{01} + \bra{10} \rho(w) \ket{10} = \frac{1 - w}{2}
\]
for a Werner state $\rho(w)$ defined in \cref{eq:werner-state}. The same result holds for $e_X$ because of the symmetry of the Werner state.
In~\cref{sec:skr with truncation}, we detail how we compute the secret-key rate with truncated waiting time distribution and Werner parameter obtained from the algorithm in \cref{sec:alg complexity}.

Since we have discrete time steps, we need an optimization algorithm which is compatible with a discrete search space.
We choose the differential evolution algorithm implemented in the SciPy-optimization library of the Python programming language \cite{storn1997differential,virtanen2020scipy}.

\section{Numerical results}
\label{sec:results}
In this section, we optimize over repeater protocols with cut-offs in order to maximize the rate at which secret key can be extracted from the produced end-to-end entanglement.
First, we use our algorithm from \cref{sec:algorithms} and the {\cutoffdifference} strategy (\cref{sec:preliminaries}) to study the effect of the cut-off on the waiting time and fidelity and show that the use of a cut-off boosts secret-key rate.
We then extend our study to two other cut-off strategies, {\cutoffmaximum} and {\cutofffidelity}, and compare their performance.
For all three cut-off strategies, we observe that the resulting repeater protocols produce secret key at significantly higher rates than their no-cut-off alternatives.
Finally, we focus on the {\cutoffdifference} strategy and analyze the sensitivity of the optimal cut-off threshold with respect to the hardware parameters.

We investigate repeater protocols with 3 nesting levels where at each nesting level the range of entanglement is doubled by an entanglement swap.
The protocol thus spans $2^3=8$ segments ($8+1=9$ nodes).
Each entanglement swap operation is preceded by a cut-off, {\it i.e.}~the scheme is of the form 
\begin{equation}
\label{protocol-form}
\textnormal{\GEN} \rightarrow (\rightarrow \textnormal{\CUTOFF} \rightarrow \textnormal{\SWAP})^3
.
\end{equation}

\begin{figure}[t]
    \centering
    \includegraphics[width=\linewidth]{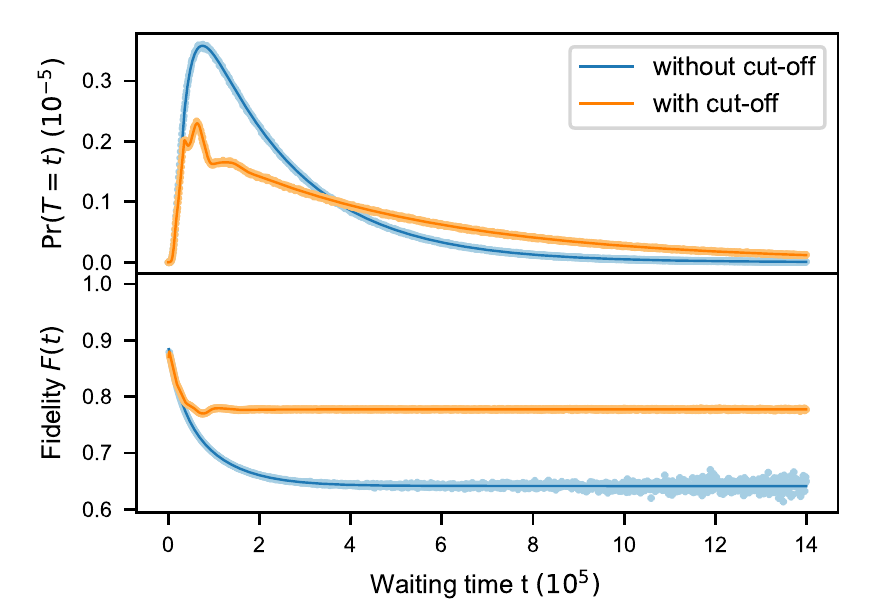}
    \caption{
The probability distribution of the waiting time $T$ and the average fidelity $F(t)$ of the end-to-end link for a protocol with and without a cut-off on entanglements' production time differences (solid lines) for a 9-node repeater protocol of the form as in~\cref{protocol-form} (unit of time is the attempt duration of elementary link generation, $L_0 / c$).
    We observe that the fidelity increases for most times $t$ while the probability that the link is produced at time $t$ shifts to larger $t$, indicating a longer waiting time.
The secret-key rates computed from the data are $0$ (without cut-off) and $0.32 \cdot 10^{-7}$ (with cut-off).
    The parameters used are $\pgen=10^{-4}$, $\pswap=0.5$, $w_0=0.98$, $\tcoh=4 \cdot 10^5$ and the cut-offs for the three nesting levels are $\tau=(1.7, 3.2, 5.5)\cdot 10^4$ (in increasing order of number of segments spanned by the {\CUTOFF} block).
Computation time $\approx$ 20 seconds for $3\cdot10^6$ time steps.
    We observe good agreement with a Monte Carlo algorithm (dots), which we use for validating the correctness of our implementation (see \cref{sec:validation} for details).
    }
    \label{fig:the waiting time and fidelity}
\end{figure}

The numerical results in this section were obtained using our open-source implementation~\cite{boxirepo} of the algorithm from \cref{sec:algorithms} on consumer-market hardware (Intel i7-8700 CPU).
We validated correctness of the implementation by comparison with an extended version of the Monte Carlo algorithm from~\cite{brand2020efficient} (see \cref{fig:the waiting time and fidelity} and~\cref{sec:validation} for details).

\begin{figure}[t]
    \includegraphics[width=\linewidth]{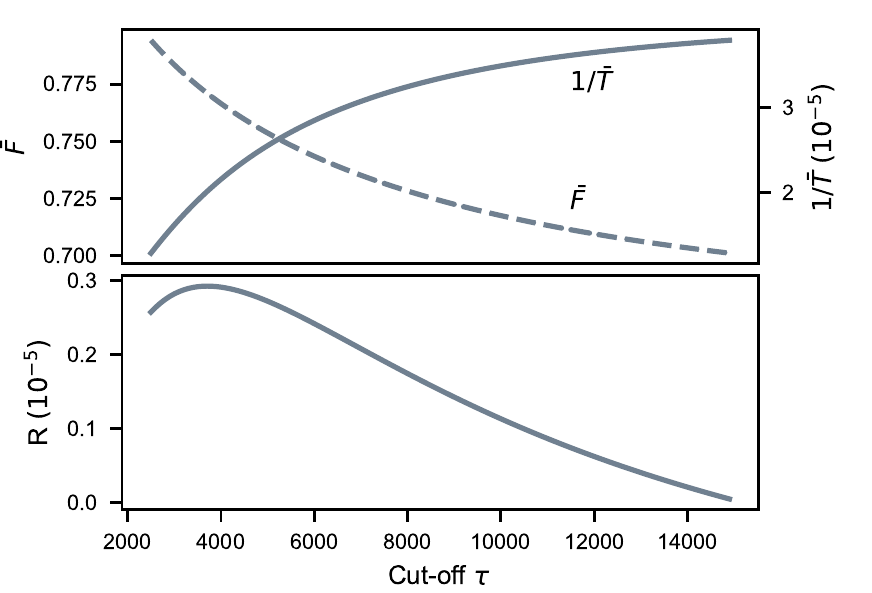}
    \caption{
    Influence of choice of cut-off on average waiting time, average fidelity and secret-key rate for repeater protocols of the form~\cref{protocol-form} where the cut-off strategy is {\cutoffdifference}.
    \textbf{(Top)} Increasing the cut-off yields higher average generation rate (reciprocal of average waiting time $\bar{T}$) but lower average fidelity $\bar{F}$.
    \textbf{(Bottom)} The secret key rate $R$ as a function of the cut-off time.
    The used parameters are $\pgen=10^{-3}$, $\pswap=0.5$, $w_0=0.98$ and $\tcoh=4\cdot10^4$.
    The chosen truncation time is $5\cdot10^5$.
    The cut-off time is chosen identical for all three swap levels.
    Unit of time is the attempt duration of elementary link generation.
    }
    \label{fig:trade-off}
\end{figure}

\subsection{Effect of DIF-TIME-CUT-OFF on the waiting time and fidelity}
\label{sec:results-difference}
We start by investigating the {\cutoffdifference} strategy, where links are discarded if their production times differ by more than a predetermined threshold $\tau$.
We compute waiting time and average fidelity for a particular choice of the cut-off threshold at each of the three levels and compare it with the protocol without cut-off (cut-off duration $\tau = \infty$ at each nesting level), see \cref{fig:the waiting time and fidelity}.
We observe that the cut-off increases fidelity at the cost of longer waiting time, as one would intuitively expect.
We further quantify the time-fidelity trade-off for a range of cut-offs in~\cref{fig:trade-off}.
For maximizing the secret key rate, we observe a single optimal choice of the cut-off threshold $\tau$.

\subsection{Extension to other cut-off strategies}

\begin{figure*}[th]
    \centering
    \includegraphics[width=\textwidth]{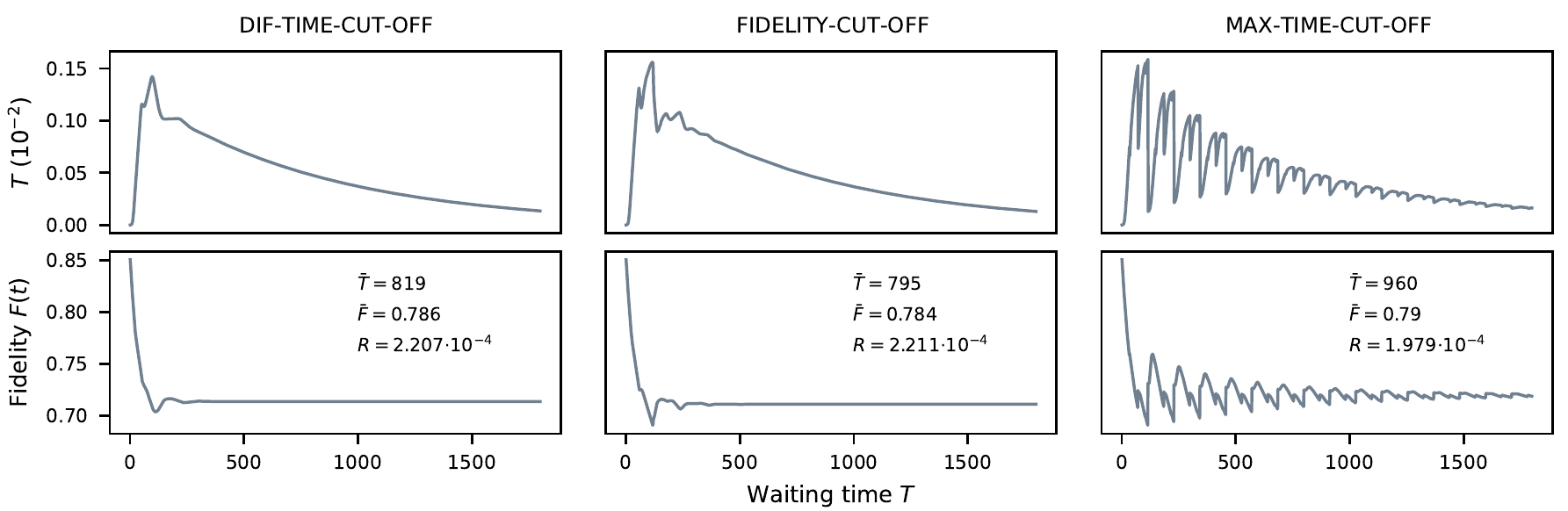}
    \caption{
        Comparison between three different cut-off strategies: cut-off on the difference of entanglements' production time ({\cutoffdifference}), the fidelity ({\cutofffidelity}) and the total waiting time ({\cutoffmaximum}, see sec.~\ref{sec:preliminaries} for definitions).
    For each strategy, we find the optimized cut-off threshold when applied to the 9-node repeater chain protocol from \eqref{protocol-form} with parameters: $\pgen=0.1$, $\pswap=0.4$, $w_0=0.98$, $\tcoh=600$.
    For each cut-off strategy, the plot shows the numerically found waiting time and fidelity distribution for the optimal protocol.
    We observe that the {\cutofffidelity} strategy yields the largest secret-key rate.
    However, the {\cutoffdifference} strategy only performs slightly worse.
    We observed the same behavior for all other parameter regimes we investigated.
    }
    \label{fig:comparison}
\end{figure*}

We extend the analysis of the previous sub-section to two other cut-off strategies: a cut-off on the fidelity ({\cutofffidelity}) and on the total waiting time ({\cutoffmaximum}, see \cref{sec:preliminaries} and \cref{tab:functions} for definitions).
To be precise, we choose the same 9-node protocol from \cref{protocol-form} and use {\cutofffidelity} and {\cutoffmaximum} as the {\CUTOFF} unit, respectively.

We observe that a single optimal cut-off threshold exists for both strategies, as we saw before already for the {\cutoffdifference} strategy in \cref{fig:trade-off}.
For each strategy, we optimize their cut-off parameters and plot the waiting time distribution and fidelity distribution in \cref{fig:comparison}.
As shown in the figure, although the {\cutofffidelity} yields the highest secret-key rate, the distribution and resulting secret-key rate of the {\cutoffdifference} strategy are very close to those of the {\cutofffidelity} strategy.
In contrast, the {\cutoffmaximum} strategy performs significantly worse in the achieved secret-key rate ($\approx 10\%$).
We find similar behavior also in other parameter regimes.

Since the {\cutoffdifference} strategy is straightforward to implement in experiments while it performs only marginally worse than the best of the three strategies ({\cutofffidelity}), we focus on this strategy for further analysis.

\subsection{Performance of the optimal cut-off for varying hardware parameters}
\begin{figure*}[t]
    \centering
    \includegraphics[width=\linewidth]{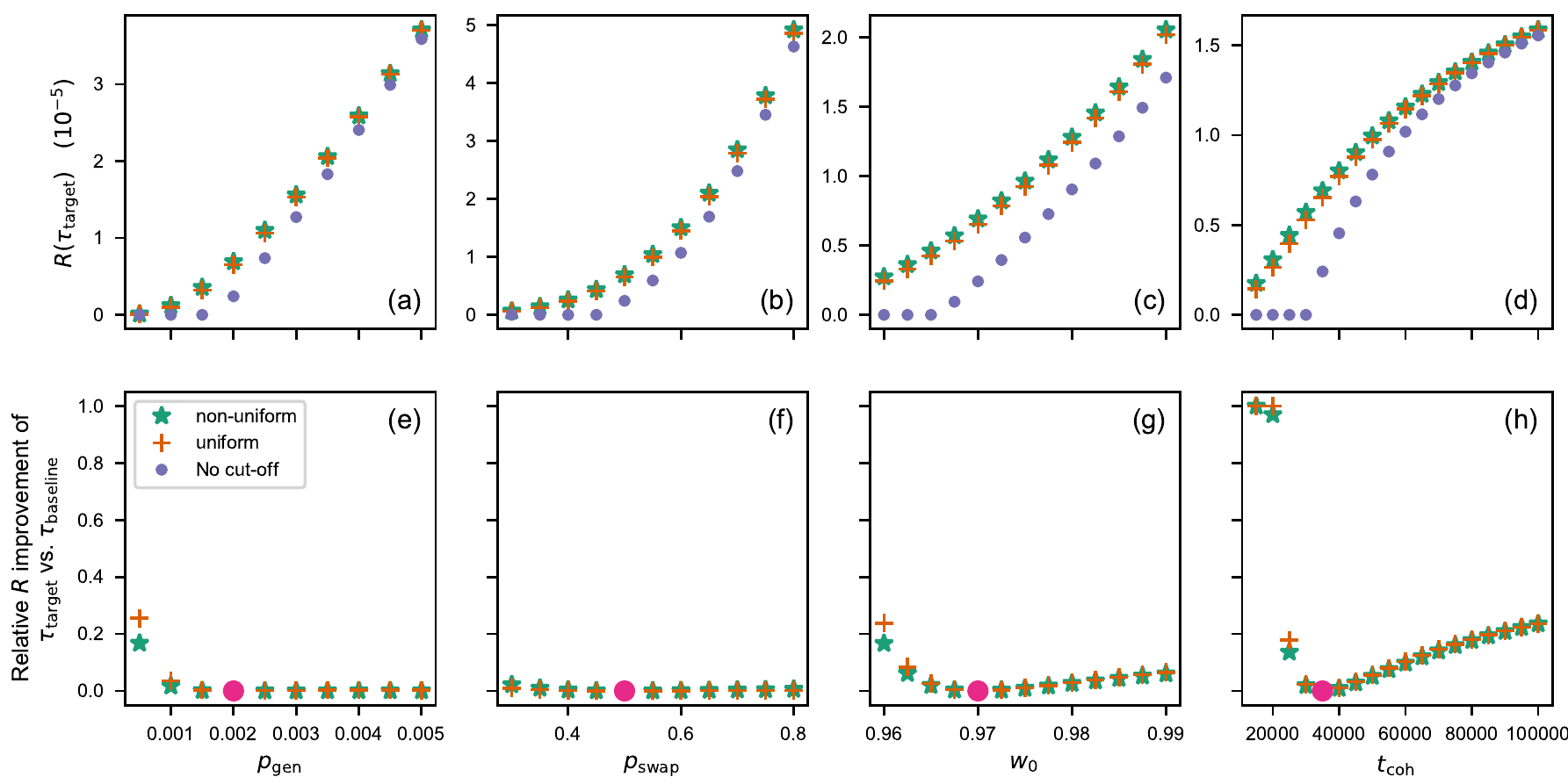}
    \caption{
        The effect of the optimal cut-off (cut-off on the difference in entanglements' production times) on secret-key rate for different hardware parameters, for the 9-node protocol as in~\cref{protocol-form}.
        We choose a set of parameters as baseline parameters ($\pgen=0.002$, $\pswap=0.5$, $w_0=0.97$ and $\tcoh=35000$) and in each plot in the figure, we vary only one of the four parameters.
        The \textbf{top plots (a-d)} show the performance of the protocol with optimized cut-offs, where the optimization is implicitly performed for each data point separately.
        The set of cut-offs we optimize over is either non-uniform (allow for different cut-offs at the three nesting levels of the protocols) or uniform (same cut-off at each level).
        We observe that the performance difference between uniform and non-uniform cut-offs is small or even negligible.
        The plots also indicate parameter regimes in which the protocol with the optimal cut-off generates key while its no-cut-off alternative does not ({\it i.e.}~the no-cut-off has zero secret-key rate).
        The \textbf{bottom plots (e-h)} show relative performance  improvement (\cref{eq:relatvice change}) of the optimal cut-off ($\tau_{\textnormal{target}}$) for a given data point, versus the optimal cut-off $\tau_{\textnormal{baseline}}$ for the baseline parameters (see above).
        The plots show that cut-off performance is most sensitive to coherence time ($\tcoh$), while it is least influenced by varying the success probability entanglement swapping ($\pswap$).
        For a detailed explanation see the main text.
        Note that the smaller the relative secret-key rate improvement (vertical axis), the closer the performance of $\tau_{\textnormal{baseline}}$ is to the performance of the optimal $\tau_{\textnormal{target}}$, which is why in the plots the best-performing `non-uniform' cut-off shows smaller relative improvement than the best-performing `uniform' cut-off.
        The purple circles refer to the baseline parameters, for which the relative improvement is 0 by definition.
    }
    \label{fig:sensitivity}
\end{figure*}

\begin{figure}[th]
    \centering
    \includegraphics[width=\linewidth]{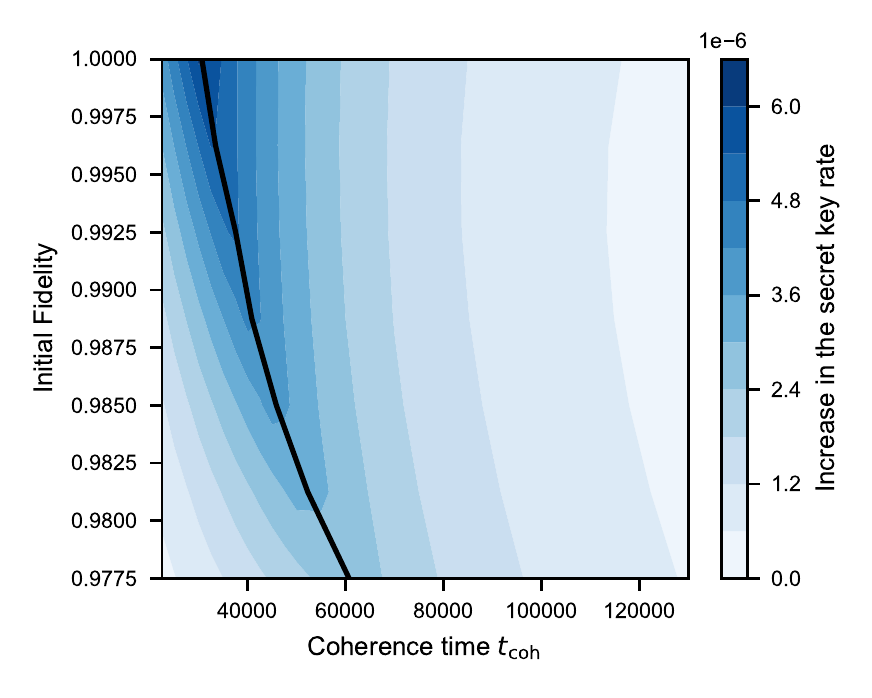}
    \caption{
The absolute increase in secret key rate with the optimal cut-off compared to no cut-off as a function of memory coherence time and fidelity of the elementary links ($=(1 + 3 w_0) / 4$, see~\cref{sec:preliminaries}), for the 9-node repeater protocols as in~\cref{protocol-form} where the used cut-off strategy is {\cutoffdifference}.
The black solid line separates the area where the no-cut-off protocol produces no secret key (left of the line) and where its secret-key rate is strictly larger than zero (right of the line).
    We observe that for the entire parameter range depicted in the figure, cut-offs increase the secret key rate and the absolute improvement is largest for parameters close to the key-producing threshold for the no-cut-off protocol (i.e.~ close to the black solid line).
    The plot consists of 126 data points on a grid and the used parameters are $\pgen=0.001$ and $\pswap=0.5$.
Time unit is the duration of a single elementary link generation attempt.
    }
    \label{fig:parameter regimes}
\end{figure}

We proceed with optimizing the cut-off in the {\cutoffdifference} strategy to maximize the secret key rate for a range of parameters.
The maximal secret-key rates for different repeater parameters are shown in~\cref{fig:sensitivity}(a-d).
We observe that cut-offs extend the parameter regime for which secret key can be generated.
To see how much one can gain in the secret key rate by using cut-offs, we choose two parameters $\tcoh$ and $w_0$ and plot the absolute increase in~\cref{fig:parameter regimes}.
We observe that the use of the optimal cut-off increases the secret key rate for the entire parameter range plotted and the improvement is largest close to the threshold parameters at which the no-cut-off protocol starts to produce nonzero secret key.

In addition, we compare uniform and non-uniform cut-offs, where `uniform' means that we choose the same cut-off time for each nesting level.
For the parameter regimes studied, we observe that non-uniform and uniform cut-off perform similarly, see~\cref{fig:sensitivity}(a-d).

Our next step is the sensitivity analysis of cut-off performance in the hardware parameters.
For this, we first choose baseline values for the four hardware parameters and find the corresponding optimal cut-off $\tau_{\textnormal{baseline}}$.
Given a target set of parameters that deviates slightly from the baseline values (optimal cut-off $\tau_{\textnormal{target}}$), we quantify the sensitivity by their relative performance difference
\begin{equation}
    \label{eq:relatvice change}
    \frac{
        R(\tau_{\textnormal{target}}) - R(\tau_{\textnormal{baseline}}) 
    }
    {
        R(\tau_{\textnormal{target}})
    }
\end{equation}
where $R$ is the secret-key rate achieved by the repeater protocol.
If this relative difference is small, the performance of cut-off is insensitive to the parameter deviation.

In~\cref{fig:sensitivity}(e-h), we plot the relative performance difference for deviations in each of the four hardware parameters separately.
We find that the performance of the baseline cut-off is influenced most by variation in coherence time, while it is largely insensitive to change in the swap success probability.
For the coherence time and the remaining two parameters, the elementary link quality and the success probability of elementary link generation, we distinguish the case where the parameter is improved and the regime where the parameter is made worse.
We observe that a worse parameter results in a significant performance difference with the optimal cutoff, while the performance difference is small when the parameter is improved.

\begin{figure}[th]
    \centering
    \includegraphics[width=\linewidth]{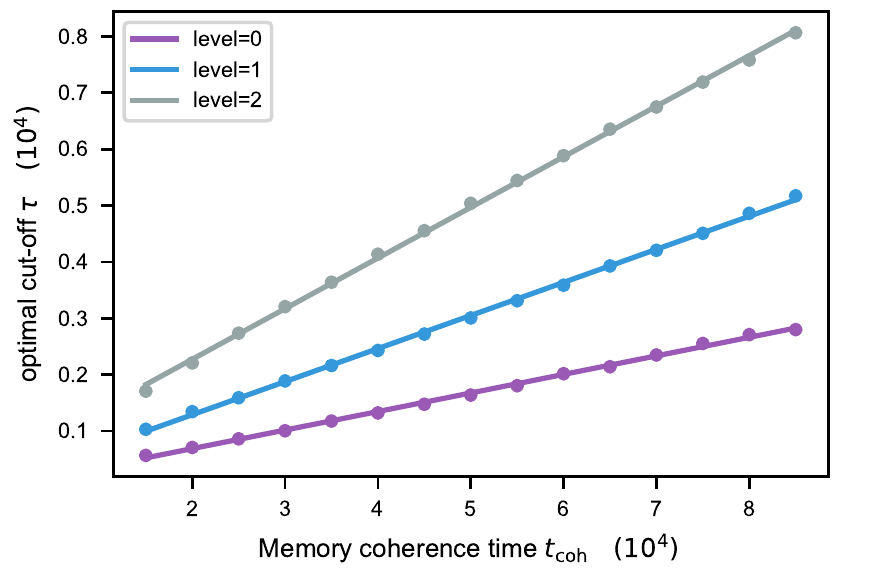}
    \caption{
        Optimal cut-off as a function of the memory coherence time in the nested 9-node repeater protocols from \cref{protocol-form}, where the cut-off strategy is ({\cutoffdifference}).
        We observe that the numerically found optimal cut-off (dots) is a linear function of the coherence time.
        Solid lines are linear fits.
        The hardware parameters used are the same as those for \cref{fig:sensitivity} (d).
        When considering the same protocol on fewer nesting levels (3 and 5 nodes, respectively), we observe similar behavior.
    }
    \label{fig:linear}
\end{figure}

We finish by investigating the most influential parameter, the coherence time, in \cref{fig:linear}. 
We observe that the optimal threshold depends approximately linearly on the memory coherence time, which could serve as a heuristic for choosing a performant cut-off.

\section{Conclusion}
\label{sec:discussion}
In this work, we optimized the secret key rate over repeater protocols including cut-offs. 
Our main tool is an algorithm for computing the probability distribution of waiting time and fidelity of the first generated end-to-end link.
The algorithm is applicable to a large class of quantum repeater schemes that can include cut-off strategies and distillation.
Its runtime is polynomial in the support size of the probability distribution of waiting time.

Our simulations show that the use of the optimal cut-off lowers the hardware quality threshold at which secret key can be generated compared to the no-cut-off alternative.
Furthermore, we observed an increase in secret-key rate for the entire regime studied for which the no-cut-off protocol produces nonzero key.

Regarding the choice of cut-off, we find that uniform cut-offs lead to a negligible reduction in the secret key rate compared to the optimal set of cut-offs which differ per nesting level.
Moreover, the optimal uniform cut-off is highly sensitive to the quality of the memory, while it is barely influenced by the success probability of swapping.
Such sensitivity could guide the heuristic cut-off optimization of more complex protocols.

\appendices
\section{Validation against a Monte Carlo algorithm}
\label[appendix]{sec:validation}
\def\sin{s_{\textnormal{in}}}
In this section, we verify that our implementation of the deterministic algorithm presented in \cref{sec:algorithms} is correct by validation against the Monte Carlo sampling algorithm from Brand et al.~\cite{brand2020efficient}.
For all repeater schemes we ran (up to \mbox{$2^{10}+1$} nodes for some parameters), we observed good agreement between the waiting time probability distribution and Werner parameter the algorithms computed, which is convincing evidence that our implementation is correct.
Fig.~\ref{fig:the waiting time and fidelity} depicts the result of a typical run.

What follows is a brief description of the Monte Carlo algorithm from Brand et al.~\cite{brand2020efficient}, including an extension to {\CUTOFF}.
Each run of the Monte Carlo algorithm samples a tuple of waiting time and Werner parameter.
It is defined recursively by having a dedicated function for each {\PUNIT} (described below) call the dedicated functions of the two {\PUNIT}s that produce its two input links.
The recursion follows the repeater protocol's tree structure (see \cref{fig:composite_protocol}), resulting in a sampling algorithm of waiting time and Werner parameter of the entire repeater protocol.

The dedicated functions for each of the four {\PUNIT}s are as follows.
If the protocol is only a {\GEN}, the Monte Carlo algorithm samples the waiting time from the geometric distribution with parameter $\pgen$ and the Werner parameter is the constant $w_0$.
For the other {\PUNIT}s, each of which takes two links as input, the algorithm begins by initializing the total elapsed time $t=0$.
Then, it enters a loop which starts by calling the dedicated functions of the {\PUNIT}s that produce the two input links, resulting in two samples $(t_A, w_A)$ and $(t_B, w_B)$.
The algorithm randomly declares `success' or `failure' according to the success probability in \cref{tab:functions}.
If it succeeds, the function breaks the loop and outputs $t + \max(t_A, t_B)$ and the resulting Werner parameter $\wout(t_A, w_A, t_B, w_B)$ (see \cref{tab:functions}).
If it fails, the total elapsed time $t$ is increased by the waiting time ($\max(t_A, t_B)$ for {\SWAP} and {\DIST}, $\min(t_A, t_B) + \tau$ for {\CUTOFF}) and the function goes back to the start of the loop.

\section{Alternative algorithm and its complexity}
\label[appendix]{sec:complexity improvement}
In \cref{sec:alg complexity}, we presented an $\mathcal{O}(\ttr^2\log{\ttr})$-algorithm for evaluating analytically-derived expressions for the waiting time distribution and average fidelity.
Here, we outline how the algorithm can be modified to achieve a complexity reduction to $\mathcal{O}(\ttr\log{\ttr})$ for protocols composed of {\PUNIT}s in \cref{tab:functions} except for {\cutofffidelity}.
Similar to the algorithm from the main text, the modified algorithm consists of two steps: first, evaluating the expressions regarding a single attempt (equations \cref{eq:semi-geom Pf,eq:semi-geom Ps,eq:semi-geom Wsuc}), followed by computing expressions regarding the whole {\PUNIT} (equations \cref{eq:fourier T,eq:fourier W}).
We show a complexity reduction for both.

For the first part, we show how to evaluate \cref{eq:semi-geom Pf,eq:semi-geom Ps,eq:semi-geom Wsuc} in time $\mathcal{O}(\ttr)$, improving on the $\mathcal{O}(\ttr^2)$ runtime of the algorithm in the main text.
Our insight here is that $p$ and $p \cdot \wout$, for {\SWAP} and {\DIST} (see \cref{tab:functions}), can always be written in the form
\begin{equation}
    \label{eq:fast iteration}
    \sum_i f^{(i)}(\tA) \cdot  g^{(i)}(\tB)
\end{equation}
where the $f^{(i)}$ and $g^{(i)}$ are arbitrary functions on the real numbers.
For instance, given $\tA \ge \tB$, we can write the success probability of distillation $\pdist$ with $f^{(1)}(\tA)=\frac{1}{2}$, $g^{(1)}(\tB)=1$ and $f^{(2)}(\tA) = \frac{1}{2} \pswap \wA(t_A) \exp(-\frac{\tA}{\tcoh})$, $g^{(2)}(\tB)=\wB(t_B) \exp(\frac{\tB}{\tcoh})$.
Consequently, each of \cref{eq:semi-geom Pf,eq:semi-geom Ps,eq:semi-geom Wsuc} can be written in the form
\begin{equation}
    \sum \limits_{\mathclap{\quad \quad \quad \quad \quad \tA,\tB: \max(\tA,\tB) = t}} 
    \Pr(\TA=\tA, \TB=\tB)
    \cdot \sum_i f^{(i)}(\tA) g^{(i)}(\tB)
    \label{eq:full-expression}
\end{equation}
which can be rewritten by splitting up the sum in the regime $t_A \geq t_B$ and $t_B > t_A$:
\begin{align}
    &&\nonumber
    \sum \limits_{\tB=0}^t
    \Pr(\TA=t, \TB=\tB)
    \cdot \sum_i f^{(i)}(t) g^{(i)}(\tB)
    \\
    &+&
    \sum \limits_{\tA=0}^{t-1}
    \Pr(\TA=\tA, \TB=t)
    \cdot \sum_i f^{(i)}(\tA) g^{(i)}(t)
    \label{eq: separable form}
    .
\end{align}
The first term in \cref{eq: separable form} can be written as
\begin{equation}
    \label{eq:genearl time iteration}
    \Pr(\TA=t)\cdot \sum_i f^{(i)}(t) \cdot G^{(i)}(t)
\end{equation}
where we have defined
\begin{equation*}
    G^{(i)}(t) = \sum_{\tB=0}^t \Pr(\TB=\tB) g^{(i)}(\tB)
    .
\end{equation*}
The expression for the second term in \cref{eq: separable form} can be found analogously.
Computing \cref{eq:genearl time iteration} for all $t$ is now performed by first computing $G^{(i)}(t)$ for all $t$, which requires linear time in $\ttr$,
and then evaluating \cref{eq:genearl time iteration} for fixed $t$ in constant time.
Therefore, the complexity for computing \cref{eq:genearl time iteration} and also for \cref{eq:full-expression} for all $t$ scales as $\mathcal{O}(\ttr)$.

This complexity holds also for protocols with {\cutoffdifference} and {\cutoffmaximum}, as the cut-off condition appears only as an additional constraint on $\tA$ and $\tB$ in the sum of \cref{eq: separable form}.
For the third cut-off strategy we consider in this work, {\cutofffidelity}, the cut-off condition is not a function of time and therefore the above method does not work.

The second part regards the evaluation of \cref{eq:semi-geom iterative convolution T,eq:semi-geom iterative convolution W} which is done exactly by the algorithm from the main text in time $\mathcal{O}(\ttr^2 \log\ttr)$.
Here, we give an $\mathcal{O}(\ttr \log\ttr)$-algorithm which evaluates the equivalent expressions in Fourier space given in \cref{sec:alg complexity} (equations \cref{eq:fourier T,eq:fourier W}) with arbitrarily small error.
We proceed in two steps.
First, we show how to evaluate the expressions in Fourier space exactly in time $\mathcal{O}(\ttr^2 \log\ttr^2)$.
Then, we show how to achieve a reduction to $\mathcal{O}(\ttr \log\ttr)$ with an arbitrarily small error.

The expressions in Fourier space (equations \cref{eq:fourier T,eq:fourier W}) hold for any $t$ in case $P_s$, $P_f$ and $\Wprep$ are defined for all $t\geq 0$.
However, in the implementation, we truncate the distribution and only have access to them for $0\leq t < \ttr$, each stored as an array of length $\ttr$, and use the discrete Fourier transform defined in \cref{eq:fourier-transform}.
The convolution defined in this way is a circular convolution:
\begin{align}
    [f_1 \tilde{\conv} f_2](t) =
    \sum_{t'=0}^t & f_1(t-t')\cdot f_2(t') +  \nonumber\\
    \sum_{t'=t+1}^{L-1} & f_1(L+t-t')\cdot f_2(t')
    \label{eq:circular convolution}
\end{align}
where $L$ is the length of the array and $\tilde{\conv}$ denotes the circular convolution.
The circular convolution introduces discrepancy compared to the linear convolution defined in \cref{eq:convolution} because
$[f_1 \tilde{\conv} f_2](t) = [f_1 \conv f_2](t) + [f_1 \conv f_2](L+t)$.
To avoid this, we pad the arrays of $P_s$, $P_f$ and $\Wprep$ with zeroes until a length of $L=\ttr^2$, which is longer than the size of $\ttr$ times convolution of arrays of size $\ttr$ (see equivalent expressions \cref{eq:semi-geom iterative convolution T,eq:semi-geom iterative convolution W}, and the algorithm presented in \cref{sec:alg complexity}).
That is, we set $P_s(t)=0$ and $P_f(t)=0$ for $\ttr \leq t < L = \ttr^2$.
With this setup, the summand in the circular convolution is always 0 for $t'>t$ and it coincides with the linear one.
The complexity of the obtained algorithm evaluating \cref{eq:fourier T,eq:fourier W} is dominated by one Fourier transform and one inverse Fourier transform on an array of length $\mathcal{O}(\ttr^2)$.
Since a Fourier transform on an array of length $L$ can be performed in time $\mathcal{O}(L\log L)$, the algorithm has a complexity of $\mathcal{O}(\ttr^2 \log \ttr^2)$.

We now show that we can reduce this complexity by zero-padding the arrays only until a length of $C\ttr$ for some predefined constant $C$, yielding an exponentially small error
\begin{equation*}
    \epsilon=\max_t\left(|\Pr(\Tout=t) - \Pr(T_{\textnormal{approx}}=t) |\right)
\end{equation*}
in $C$ of the distribution $\Pr(T_{\textnormal{approx}}=t)$ obtained with circular convolution.
The resulting algorithm has complexity of $\mathcal{O}(C\ttr \log (C\ttr)) = \mathcal{O}(\ttr \log \ttr)$.

The motivation behind this reduction is that $\Pr(\Tout=t)$ is the sum of all possible sequences of failed attempts (see \cref{eq:semi-geom iterative convolution T}) and is exponentially decreasing for large $t$.
For a fixed number of attempts $k$, the probability results from a successful attempt after at least $k-1$ failed attempts.
Therefore, it has an occurrence probability of at most $(1-p)^{k-1}$, where $p$ is the success probability for a \PUNIT.
To see this mathematically, we use the Young's convolution inequality \cite{bogachev2007measure} and obtain
\begin{equation*}
    \left\lVert \Conv\limits_{j=1}^{k-1} \Pf^{(j)} \conv \Ps \right \rVert 
    \le
    \left\lVert \Pf \right\rVert^{k-1}  \left\lVert \Ps \right \rVert
    \le
    (1-p)^{k-1}
\end{equation*}
where the norm is defined by $\left\lVert f(t) \right\rVert = \sum_t f(t)$.
In addition, note that
\begin{equation*}
    \left[\Conv\limits_{j=1}^{k-1} \Pf^{(j)} \conv \Ps \right] (t)= 0 \quad \textnormal{for} \quad t \ge k\ttr
\end{equation*}
because $\Pf(t)$ and $\Ps(t)$ are finite arrays of length $\ttr$.
Hence, for $t \ge K\ttr$, we only need to consider the terms with $k\geq K+1$, {\it i.e.}~cases with at least $K$ failed attempts.
As a result, we obtain a bound for the probability given in \cref{eq:semi-geom iterative convolution T} for $t \ge K\ttr$:
\begin{equation*}
    \Pr(\Tout=t) \le \sum_{k=K+1}^{\infty}(1-p)^{k-1} = \frac{(1 - p)^{K}}{p}
    .
\end{equation*}

The above expression bounds the distribution with an exponentially decreasing probability with respect to the minimal number of failed attempts, which we now use to bound the error.
Because of the circular convolution \cref{eq:circular convolution}, if we only zero-pad to $C\ttr$, the obtained distribution is given by
\begin{equation*}
    \Pr(T_{\textnormal{approx}}=t) = 
    \sum_{j=0}^{\infty}
    \Pr(\Tout=t + jC\ttr)
\end{equation*}
for $0\leq t < C\ttr$.
That is, the probability for $t>C\ttr$ ($j>0$) will be added to the first $C\ttr$ elements, introducing an error in the final result.
This error is bounded by
\begin{equation*}
    \epsilon =
    \sum_{j=1}^{\infty} 
    \frac{(1 - p)^{jC}}{p}
    \le \frac{(1 - p)^{C}}{p^2}
    ,
\end{equation*}
which is exponentially small in $C$.
The same bound can be given in analog for the calculation of $\Wout(t)$ defined in \cref{eq:semi-geom iterative convolution W} by noticing that $\Wprep(t) \leq 1$.

The above bound is only for a single {\PUNIT} and does not account for the propagation of noise among different levels.
However, in practice, as long as one chooses a $C$ large enough so that the error on each array value is below the numerical accuracy, this improved algorithm gives the same result as the algorithm provided in the main text.
In addition, the above bound is very loose.
In our numerical study, we find that, if the truncation time $\ttr$ is chosen so that more than 99\% distribution is covered, it suffices to triple the size of the array during the calculation, {\it i.e.}~set $C=3$.

\begin{figure}[t]
    \centering
    \includegraphics[width=\linewidth]{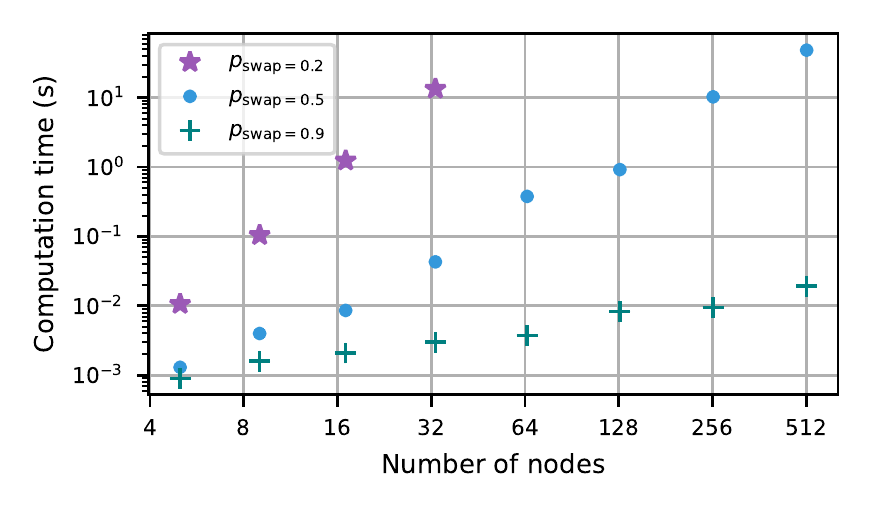}
    \caption{
    Computation time of the algorithm from \cref{sec:complexity improvement} as a function of the number of nodes in the repeater chain using consumer-market hardware (Intel i7-8700 CPU).
    We plot the computation time for three different $\pswap$ and for protocols of the form
    $\textnormal{\GEN} \rightarrow (\rightarrow \textnormal{\CUTOFF} \rightarrow \textnormal{\SWAP})^n$,
    similar to \cref{protocol-form}, where $n$ is the nesting level and the number of nodes is $2^n+1$.
    The truncation time is chosen such, that 99\% of the probability mass is covered.
    Note that the plot's axes are both given in logarithmic scale; in such a log-log plot, a polynomial function is represented as a line.
    The used cut-off strategy is {\cutoffdifference} and the other parameters used are: $\pgen=0.1$, $w_0=1.0$, $\tcoh=500/\pswap^{n-1}$, $\tau=42/\pswap^{n-1}$.
    In this plot, the number of truncation time steps goes up to about $10^6$.
    \label{fig:computation time}
    }
\end{figure}

Although in general there exists no efficient algorithm which captures a constant fraction of the probability mass for protocols including a cut-off (see \cref{sec:alg complexity}), we numerically find that the algorithm outlined above scales polynomially in the number of nodes in some parameter regimes, see \cref{fig:computation time}.

\section{Calculation of the secret-key rate}
\label{sec:skr with truncation}
Here, we show how we calculate the secret-key rate with truncated waiting time distribution.

One could think of the secret-key rate, computed with finite truncation time $\ttr<\infty$, as an approximation of the real secret-key rate or, alternatively, as the rate achieved by the following repeater protocol.
The protocol starts with the two parties at the end nodes agree on a truncation time $\ttr$.
If up to $t=\ttr$ the end-to-end link has not been delivered, the protocol terminates and restarts from {\GEN}.
Therefore, the number of protocol executions follows the geometric distribution with success probability $p_{\textnormal{tr}}=\Pr(T\leq \ttr)$.
The waiting time for a failed protocol is $\ttr$ while for a successful one it follows the waiting time distribution $\Pr(T=t)$ for $t<\ttr$.
The average total waiting time is then the sum of the time consumed in failed and successful executions:
\begin{equation*}
    \bar{T} =
    \ttr \cdot \left(\sum_{k=1}^{\infty} k \cdot p_{\textnormal{tr}} (1-p_{\textnormal{tr}})^{k}\right)
    +
    \frac
    {
        \sum_{t=1}^{\ttr} t \cdot \Pr(T=t)
    }
    {
        \Pr(T \le \ttr)
    }
    .
\end{equation*}
Accordingly, the average Werner parameter is an average over the successful execution
\begin{equation*}
    \bar{W} = \frac{\sum_{t=1}^{\ttr} W(t) \cdot \Pr(T=t)}{\Pr(T \le \ttr)}.
\end{equation*}
With the above equations, we calculate the secret-key rate defined in \cref{eq:secret key rate}.
In this work, we choose heuristically a $\ttr$ such that $\Pr(T\leq\ttr) \ge 99\%$.
With this choice, the difference in the secret key rate between protocols with finite and infinite $\ttr$ is negligibly small.

\section*{Acknowledgment}
The authors would like to thank Sebastiaan Brand, Kenneth Goodenough and Filip Rozp\k{e}dek for helpful discussions.
This work was supported by the QIA project (funded by European Union's Horizon 2020, Grant Agreement No. 820445) and by the Netherlands Organization for Scientific Research (NWO/OCW), as part of the Quantum Software Consortium program (project number 024.003.037 / 3368).
Boxi Li was supported by the IDEA League student grant programme.

\bibliographystyle{IEEEtran}

\end{document}